\documentclass[prd,twocolumn,superscriptaddress,floatfix,nofootinbib]{revtex4-2}
\pdfoutput=1 
\usepackage[T1]{fontenc} 
\usepackage{amsmath,amssymb,amsfonts,amsthm}
\usepackage{graphicx}
\usepackage[usenames,dvipsnames]{color}
\usepackage{enumitem}
\usepackage{verbatim}
\usepackage[percent]{overpic}
\usepackage{rotating}
\usepackage{hyperref}
\usepackage{array}
\usepackage{mathtools}
\usepackage{lmodern}
\usepackage[normalem]{ulem}
\usepackage{braket}
\usepackage{tensor}
\usepackage{dsfont}
\usepackage{bm}
\usepackage{tikz}

\usepackage{mathrsfs}

\usetikzlibrary{decorations.pathmorphing}
\usepackage{CJKutf8}

\newcommand{\comm}{\left[\partial_{\tau\ts{A}}\hat\phi(\sx\ts{A}(\tau\ts{A})),\partial_{\tau\ts{B}}\hat\phi(\sx\ts{B}(\tau\ts{B}))\right]}

\usetikzlibrary{positioning}
\usetikzlibrary{calc,through,backgrounds}

\theoremstyle{definition}
\newtheorem{definition}{Definition}[section]
\newtheorem{theorem}{Theorem}[section]

\newcommand{\kako}[1]{\left( #1 \right)}
\newcommand{\kagikako}[1]{\left[ #1 \right]}
\newcommand{\ts}[1]{ _{\text{#1}} }

\allowdisplaybreaks

\newcommand{\eri}[1]{\leavevmode{\color{OliveGreen}{\textbf{[Erickson:} #1]}}}

\DeclareMathOperator{\Tr}{Tr}

\newcommand{\dd}{\text{d}}

\newcommand{\rr}[1]{\left(#1\right)}

\newcommand{\sx}{\mathsf{x}}

\newcommand{\ii}{\mathsf{i}}
\newcommand{\pd}{\partial}

\begin{document}

\title{Harvesting Entanglement
with  Detectors Freely Falling into a Black Hole}

\author{Kensuke Gallock-Yoshimura}
\email{kgallock@uwaterloo.ca}
\affiliation{Department of Physics and Astronomy, University of Waterloo, Waterloo, Ontario, N2L 3G1, Canada}

\author{Erickson Tjoa}
\email{e2tjoa@uwaterloo.ca}
\affiliation{Department of Physics and Astronomy, University of Waterloo, Waterloo, Ontario, N2L 3G1, Canada}
\affiliation{Institute for Quantum Computing, University of Waterloo, Waterloo, Ontario, N2L 3G1, Canada}

\author{Robert B. Mann}
\email{rbmann@uwaterloo.ca}
\affiliation{Department of Physics and Astronomy, University of Waterloo, Waterloo, Ontario, N2L 3G1, Canada}
\affiliation{Perimeter Institute for Theoretical Physics,  Waterloo, Ontario, N2L 2Y5, Canada}

\begin{abstract}
We carry out the first investigation of  the entanglement and mutual information harvesting protocols  for detectors freely falling into a black hole. Working in $(1+1)$-dimensional Schwarzschild black hole spacetime, we consider two pointlike Unruh-DeWitt (UDW) detectors in different combinations of free-falling and static trajectories. Employing a generalization of relative velocity suitable for curved spacetimes, we find that the amount of correlations extracted from the black hole vacuum, at least outside the near-horizon regime, is largely kinematic in origin (i.e. it is mostly due to the relative velocities of the detectors). Second, correlations can be harvested purely from the black hole vacuum even when the detectors are causally disconnected by the event horizon. Finally, we show that the previously known `entanglement shadow' near the horizon is indeed absent for the case of two free-falling-detectors,  since their relative gravitational redshift  remains finite as the horizon is crossed, in accordance with the equivalence principle.

\end{abstract}

\maketitle
\flushbottom

\section{Introduction}

For the past few decades, entanglement has been viewed as a resource of quantum information processing \cite{nielsen2000quantum,chitambar2019quantum,Brandao2015reversibleQRT}, and its role in quantum field theory (QFT) and gravity theory has attracted increasing attention. For example, entanglement theory is often used in the black hole information problem \cite{hawking1975particle,almheiri2013black,marolf2017black,haco2018black}, where entanglement is treated as an intermediary tool for unraveling the quantum nature of gravity. 

The fact that vacuum states in QFT are highly entangled was first studied formally in \cite{summers1985bell,summers1987bell}, followed by the operational formulation by Valentini \cite{Valentini1991nonlocalcorr} and Reznik \textit{et al}. \cite{reznik2003entanglement, reznik2005violating}. It was found that an uncorrelated pair of atoms can extract entanglement from the vacuum of the quantum fields. This process is more recently known as entanglement harvesting. As a quantum information protocol, a more simplified model of the atom-field interaction known as the Unruh-DeWitt (UDW) detector model \cite{Unruh1979evaporation, DeWitt1979} is widely used to understand the underlying the essential physics so long as no angular momentum is exchanged.  Recent extensive research on the entanglement harvesting protocol has shown that it is sensitive to the detectors' motion (e.g. acceleration \cite{salton2015acceleration}), as well as the properties of the background geometry such as spacetime dimension \cite{pozas2015harvesting}, curvature, black holes \cite{Steeg2009,kukita2017harvesting,henderson2018harvesting,ng2018AdS,Tjoa2020vaidya,robbins2020entanglement}, causal structure 
\cite{Henderson2020temporal},
topology \cite{smith2016topology}, and boundary conditions \cite{henderson2019entangling,cong2019entanglement,cong2020horizon}. 

In this article, we are particularly interested in   entanglement harvesting in black hole spacetimes using inertial, free-falling detectors.  So far, the harvesting protocol in a (rotating) BTZ black hole \cite{henderson2018harvesting,robbins2020entanglement}, Schwarzschild and Vaidya spacetimes \cite{Tjoa2020vaidya} have been studied using two static detectors. In both cases, it was found that static detectors are unable to extract entanglement from the vacuum when they are close to an event horizon. This entanglement shadow (or `death zone') {appears to be a characteristic feature} of entanglement harvesting in black hole spacetimes.  However, extraction of entanglement from the black hole interior has not yet been investigated.  What role the horizon plays in this regard is not obvious. Furthermore, since the origin of the entanglement shadow is often attributed to divergent gravitational redshift as the detectors are placed closer to the horizon, there is warrant for seeing whether or not this occurs for inertial trajectories.  
 
In this paper  we address these two questions by considering correlation harvesting protocols between two pointlike two-level detectors when one or both of them freely fall toward a (1+1) dimensionally reduced Schwarzschild black hole. This lower-dimensional setting introduces considerable simplification insofar as complicated sums over field modes are avoided (see e.g. \cite{Ng2014Schwarzschild,Ng2018newtechniques}). Despite this, the setting is notably more complicated than in previous studies, particularly in terms of numerical evaluation of the bipartite density matrix of the detectors, because the free-falling trajectory has a time-dependent gravitational redshift. To this end, we employ the derivative coupling variant of the Unruh-DeWitt model \cite{Aubry2014derivative,Aubry2018Vaidya,Tjoa2020vaidya} in order to remove both infrared (IR) ambiguities  and the lack of Hadamard short-distance property associated with massless scalar fields in two-dimensional spacetimes that plague the usual linear amplitude coupling. Our approach also allows us to investigate for the first time harvesting in the black hole interior.

 We present three main results. First, using a generalization of relative velocity suitable for curved spacetimes   \cite{bolos2007intrinsic}, we show that the amount of correlations obtained from the black hole vacuum, at least outside the near-horizon regime, is largely kinematic in origin. In other words,  it is mostly due to the effective relative velocities of the detectors rather than  intrinsic properties of the gravitational field. From this analysis, we find that when one of the detectors freely falls (starting at rest from infinity), in general less entanglement can be harvested than the case when both detectors are static. Thus we identify relative velocity (and acceleration) as the main source of degradation for entanglement harvesting. The kinematic nature of this effect implies that the same is true in flat space. This suggests that any intrinsic contribution from the gravitational field that cannot be accounted for this way is necessarily confined to the near-horizon regime or black hole interior.  We demonstrate this by comparing the scenario where Alice is free-falling and Bob is static in the Boulware vacuum (where relative acceleration is negligible) to the case in flat space where their relative velocity is the same. 
 
Second, we show that while in general Alice's free-falling motion (keeping Bob static) tends to lead to lower correlations, they can still harvest correlations even when they are causally disconnected by the horizon (i.e. Alice is in the black hole interior). When both detectors are free-falling, we can show using a signalling estimator defined in \cite{Causality2015Eduardo,Tjoa2020vaidya} that the increase in efficiency of the harvesting protocol is in some sense due to increasing assistance from communication mediated by the field.
 
Third, we find that for two free-falling detectors  the entanglement shadow is indeed absent. This is in accord with the equivalence principle and can be attributed to the fact that the trajectories are inertial so that the relative gravitational redshift remains finite during horizon-crossing. This is in contrast to static detectors, which cannot maintain static trajectories at the horizon, and is manifest in the form of increasing local noise as the horizon is approached, which effectively cuts off all correlations. 

Our paper is organized as follows. In Section~\ref{sec: geometry} we review the construction of a quantum massless scalar field in Schwarzschild background and the associated coordinate systems adapted to both static and free-falling observers. In Section~\ref{sec: UDW-model} we describe the derivative-coupling variant of the Unruh-DeWitt particle detector model and review the notion of signalling estimator for analyzing causal relations between the two detectors. In Section~\ref{sec: results} we describe our main results in full detail, and we conclude with some future directions in Section~\ref{sec: conclusion}.

In this paper we use   natural units $c = \hbar = 1$. 
We take the metric $g$ to be such that $g(\mathsf{V},\mathsf{V})=g_{\mu\nu}V^\mu V^\nu <0$ if $\mathsf{V} = V^\mu\partial_\mu$ is a timelike vector, since the metric signature is ambiguous in two dimensions. We also use the shorthand $\sx\equiv x^\mu$ to denote the spacetime events whose coordinates are given by $x^\mu$.

\section{Klein-Gordon field in Schwarzschild spacetime}
\label{sec: geometry}


In the next two sections we first review the geometrical and quantum field-theoretic aspects of a quantum massless scalar field in a Schwarzschild background spacetime, following discussion in \cite{Aubry2018Vaidya,Tjoa2020vaidya}. We will review three coordinate systems that are naturally associated with the three standard vacua --- Boulware, Unruh, and Hartle-Hawking states --- and also a coordinate system adapted to a class of free-falling observers.

\subsection{Schwarzschild geometry}

Consider a $(3+1)$-dimensional Schwarzschild spacetime described by the metric
\begin{align}
    \dd s^2
    &= -f(r) \dd t\ts{s}^2 + \dfrac{ 1}{ f(r) }\dd r^2  + r^2 (\dd \theta^2 + \sin^2\theta \dd \varphi^2)\,, \label{eq: Schw-coords} \\
    f(r) &= 1-\frac{r\ts{s}}{r}\,,
\end{align}
where $r\ts{s}=2G\mathsf M$ is the Schwarzschild radius and $\mathsf M\geq 0$ is the ADM mass. It is convenient to write $M=G\mathsf{M}$ so that $r\ts s$ has units of length. The standard Schwarzschild coordinates are given by $(t\ts{s},r,\theta,\phi)$, where the subscript `S' will be useful because we will later be considering another coordinate system. For static spherically symmetric black holes, this metric is valid only for $r>r\ts{s}$ due to the coordinate singularity at $r = r\ts{s}$. 
The null hypersurface $r=r\ts{s}$ defines the event horizon of the black hole.



\begin{figure*}[tp]
    \centering
    \includegraphics[width=13cm]{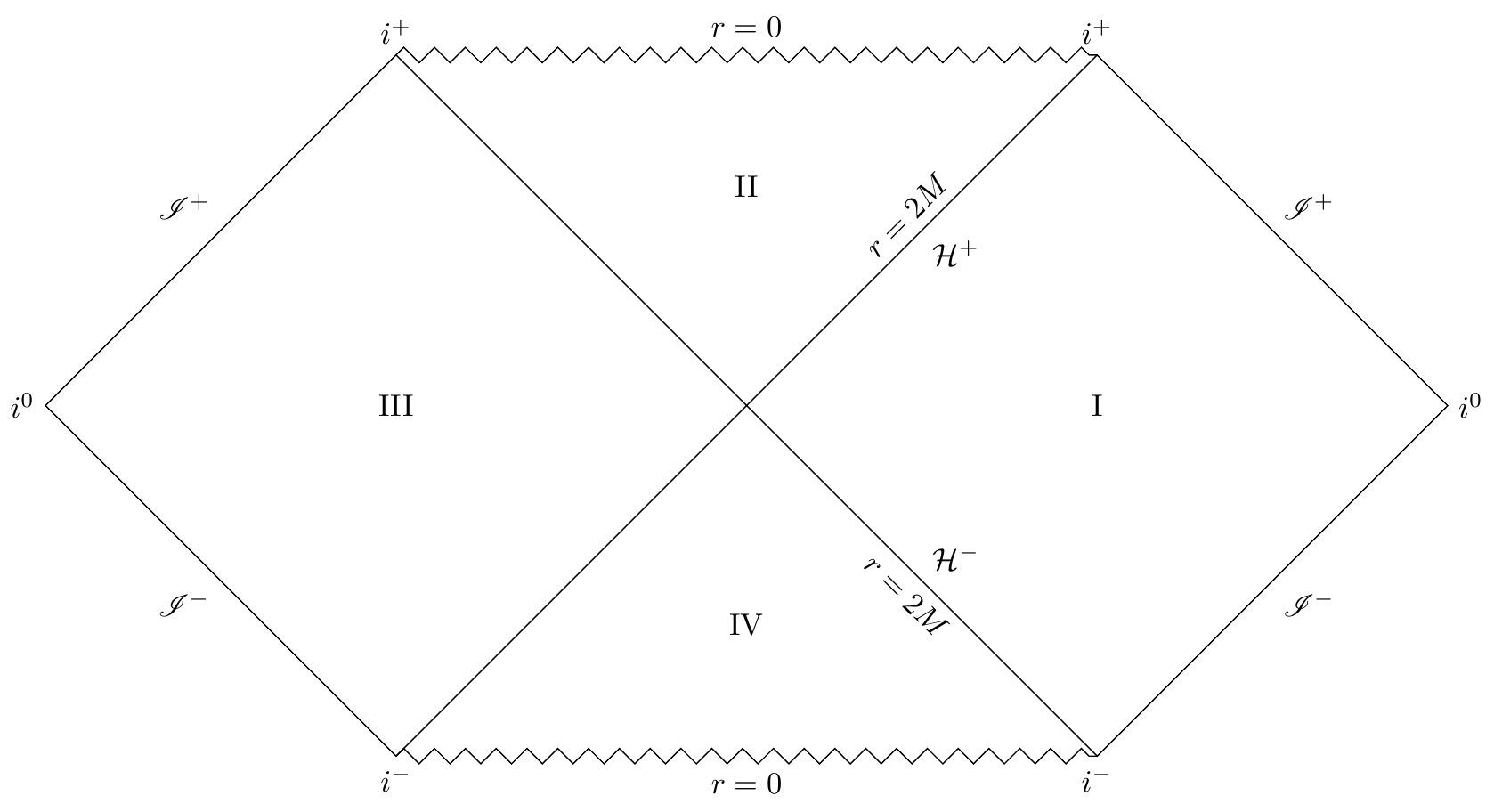}\\
    \caption{\,Conformal diagram for Schwarzschild spacetime, following \cite{Tjoa2020vaidya}.}
    \label{fig: Penrose1}
\end{figure*}

We can extend the coordinate system by first introducing the tortoise coordinate $r_\star$ defined by
\begin{align}
    r_{\star}\coloneqq r+r\ts{s} \ln \left| \frac{r}{r\ts{s}}-1 \right|\,,
\end{align}
and then defining the null coordinates $v\coloneqq t\ts{s}+r_{\star},~u\coloneqq t\ts{s}-r_{\star}$. With this, the metric now reads
\begin{align}
    \dd s^2 &= -\frac{r\ts{s}}{r}e^{-\frac{r}{r\ts{s}}}e^{\frac{v-u}{2r\ts{s}}}\dd u\,\dd v +r^2\rr{\dd \theta^2+\sin^2\theta\dd\varphi^2}\,.
\end{align}
Finally, introducing new coordinates
\begin{align}
    U\coloneqq -2r\ts{s} e^{ -u/2r\ts{s} },~
    V\coloneqq 2r\ts{s} e^{ v/2r\ts{s} }\,,
\end{align}
the extension to region II in Fig. \ref{fig: Penrose1} is obtained by considering the coordinate system $(U,v,\theta,\phi)$ where $U,v\in \mathbb{R}$ and the metric reads
\begin{align}
    \dd s^2 = -\frac{2r\ts{s}^2}{r}e^{-\frac{r}{r\ts{s}}+\frac{v}{2r\ts{s}}}\dd U \dd v + r^2\rr{\dd \theta^2+\sin^2\theta\dd\phi^2}\,.
    \label{eq: EF-coords}
\end{align}
Note that here $r$ is an implicit function of $U$ and $v$. 
The maximal analytic extension is obtained by considering the coordinate system $(U,V,\theta,\phi)$ where $U,V\in\mathbb R$ and the metric reads 
\begin{align}
    \dd s^2=
    -\dfrac{ r\ts{s} }{r} e^{ -r/r\ts{s} } \dd V \dd U 
    + r^2 \rr{\dd \theta^2 + \sin^2\theta \dd \varphi^2}\,. 
    \label{eq: Kruskal-coords}
\end{align}
Thus we have obtained three distinct coordinate systems for the Schwarzschild black hole spacetime: \textit{Schwarzschild coordinates} with metric \eqref{eq: Schw-coords}, \textit{Eddington-Finkelstein coordinates}\footnote{Strictly speaking Eddington-Finkelstein coordinates refer to coordinates ($u,r,\theta,\varphi$) or $(v,r,\theta,\varphi)$, but we will borrow this name because they share the same region of validity (regions I and II) without any analytic extension.} with metric \eqref{eq: EF-coords}, and \textit{Kruskal-Szekeres coordinates} with metric \eqref{eq: Kruskal-coords}. 
These three coordinate systems are naturally adapted for definitions of the three standard vacuum states of quantum fields in this background spacetime, as we will see in the next subsection.

Finally, in this paper we will consider a class of free-falling observers that are infalling from infinity, possibly towards the curvature singularity at $r=0$. For this purpose, it will not be sufficient for us to simply solve for radial geodesics in Schwarzschild coordinates because the coordinate systems do not apply for free-fallers inside the horizon. The coordinate system we need for this class of observers that is also regular at the event horizon $r=r\ts{s}$ is the Painlev\'e-Gullstrand (PG) coordinate system (see \cite{Martel2001generalPG} and references therein), which is constructed based on a free-falling observer's proper time. 

The PG coordinate system is adapted to free-falling observers starting at rest at spatial infinity, with metric given by 
\begin{align}
   \dd s^2 &= - f(r) \dd t\ts{PG}^2 + 2 \sqrt{ 1-f(r) } \dd t\ts{PG} \dd r + \dd r^2 \notag \\
   &\hspace{5mm}+ r^2 (\dd \theta^2 + \sin^2\theta \dd \varphi^2),
\end{align}
where $t\ts{PG}$ is the PG (coordinate) time. The PG coordinates $(t_{\text{PG}},r,\theta,\phi)$ for infalling observers are obtained from the Schwarzschild metric \eqref{eq: Schw-coords} by using the coordinate transformation 
\begin{align}
    t\ts{PG}&=t+2 r\ts{s}\kako{\sqrt{ \dfrac{r}{r\ts{s}} }+\dfrac{1}{2}\ln\left|
            \dfrac{ \sqrt{ r/r\ts{s} } -1 }{ \sqrt{ r/r\ts{s} } + 1 }\right|}\,,
\end{align} 
Time reparametrization invariance allows us to fix $t\ts{PG}=0$ at the singularity $r=0$ and so $t\ts{PG}<0$ for all $r>0$. A remarkable property of the PG coordinates is that the induced metric at constant $t\ts{PG}$ slices are flat; thus proper distances between two fixed radial coordinates $r_1,r_2$ will be given simply by $\Delta r = |r_2-r_1|$. 

In what follows, we will consider the (1+1)-dimensional reduction of the Schwarzschild spacetime, by truncating the angular part. While we will lose the physics that depends on angular variables such as the graybody factors due to the gravitational potential (associated with spherical harmonic parts of the wave equation) and the physics associated with orbital motion, much of the essential features of quantum field theory in curved spacetimes will remain. For example, the detailed balance condition associated with detector thermalization in the Hartle-Hawking state  can be obtained \cite{Aubry2018Vaidya,Tjoa2020vaidya}. This dimensional reduction allows us to borrow conformal techniques and obtain closed-form expressions for the two-point functions of the quantum field, thus simplifying the setup considerably.

\subsection{Klein-Gordon field and vacuum two-point functions}
Let $\phi(\sx)$ be a real-valued massless Klein-Gordon field in $(1+1)$-dimensional Schwarzschild spacetime. 
The Klein-Gordon equation is given by 
\begin{align}
    \frac{1}{\sqrt{-g}}\partial_\mu \rr{\sqrt{-g}g^{\mu\nu}\partial_\nu}\phi =0 ,
    \label{eq: KGE}
\end{align}
where $g\coloneqq\det g_{\mu \nu}$ is the metric determinant. After canonical quantization, the quantum field admits Fourier mode decomposition of the form
\begin{align}
    \hat \phi(\sx) = \int \dd k\, \rr{\hat a_k^{\phantom{\dagger}}u^{\phantom{*}}_k(\sx) +\hat a_k^{\dagger}u_k^*(\sx) }\,.
\end{align}
The mode  (eigen)functions $\{u_k(\sx)\}$ satisfy the orthogonality conditions
\begin{align}
    &(u_{k},u_{k'})=\delta(k-k')\,,~(u_{k}^*,u_{k'}^*)=-\delta(k-k')\,, \notag \\
    &(u_k^{\phantom{*}},u^*_{k'})=0\,,
\end{align}
where $(f,g)$ is the Klein-Gordon inner product of $f,g$ given by
\begin{align}
(f,g) = -\ii\int_{\Sigma}\dd \Sigma^\mu \sqrt{-g}\rr{f\nabla_\mu g^* - g^*\nabla_\mu f } 
\end{align}
with respect to the Cauchy surface $\Sigma$.

The definition of a vacuum state of the field depends on the choice of timelike Killing vector field with respect to which the positive frequency modes $u_k(\sx)$ are defined \cite{birrell1984quantum,Aubry2018Vaidya,Tjoa2020vaidya}.  
There are three standard choices of vacuum states that are unitarily inequivalent and are associated with different regions of spacetime:
\begin{enumerate}[nolistsep,leftmargin=*,label=(\alph*)]
    \item Boulware vacuum $\ket{0\ts{B}}$, with $u_k(\sx)$ being positive frequency with respect to $\partial_t$. The state is defined on exterior region I of Fig.~\ref{fig: Penrose1}.
    \item Unruh vacuum $\ket{0\ts{U}}$, with $u_k(\sx)$ being positive frequency  with respect to $\partial_U$ on $\mathcal{H}^-$ and $\partial_v$ on $\mathscr{I}^-$. The state is defined on regions I and II of Fig. \ref{fig: Penrose1}.
    \item Hartle-Hawking vacuum $\ket{0\ts{H}}$, with $u_k(\sx)$ being positive frequency  with respect to horizon generators $\partial_U$ and $\partial_V$. The state is defined on the full maximally extended Schwarzschild spacetime regions I-IV of Fig. \ref{fig: Penrose1}.
\end{enumerate}
The Boulware vacuum is the vacuum state that reproduces the Minkowski vacuum in the large $r$ limit, whereas the Hartle-Hawking vacuum is the vacuum state that reproduces a thermal state in flat space in the large $r$ limit. The
Unruh vacuum is, by construction, one that mimics radiation outflux, effectively by replacing the ingoing Hartle-Hawking modes with ingoing Boulware modes.

In terms of field observables, the distinct vacua (denoted $\ket{0_\alpha}$ where $ \alpha=\text{B, U, H}$) can be specified by the vacuum Wightman two-point functions:
\begin{align}
    W_\alpha(\sx,\sx') \coloneqq \Tr\rr{\hat\phi(\sx)\hat\phi(\sx')\ket{0_\alpha}\!\bra{0_\alpha}}\,,
    \label{eq: wightman-general}
\end{align}
and all higher $n$-point vacuum correlation functions can be obtained as products of the vacuum two-point functions. For each vacuum state (here $\alpha=\text{B, U, H}$) we have
\begin{subequations}
\begin{align}
    W\ts{B}(\sx,\sx') &= -\frac{1}{4\pi}\log\left[-\Lambda^2(\Delta u-\ii\epsilon)(\Delta v-\ii\epsilon)\right]\,,
    \label{eq: wightman-boulware}\\
    W\ts{U}(\sx,\sx') &= -\frac{1}{4\pi}\log\left[-\Lambda^2(\Delta  U-\ii\epsilon)(\Delta v-\ii\epsilon)\right]\,,
    \label{eq: wightman-eddington}\\
    W\ts{H}(\sx,\sx') &= -\frac{1}{4\pi}\log\left[-\Lambda^2(\Delta  U-\ii\epsilon)(\Delta  V-\ii\epsilon)\right]\,,
    \label{eq: wightman-hartlehawking}
\end{align}
\end{subequations}
where $\Lambda>0$ is an IR cutoff inherent in (1+1) massless scalar field theory. 

We make a parenthetical remark that in principle, one could try to perform canonical quantization with respect to the PG coordinates where the vacuum state (which we may call PG vacuum $\ket{0\ts{PG}}$) is associated with a freely falling observer (see e.g.,  \cite{WEINSTEIN2002mirrors,Melnikov2004free-faller,Goto2019free-faller} for related discussions). This will be slightly more involved due to the cross-term in the metric. However by construction this state will be regular across the horizon and is well-defined on regions I and II of the Schwarzschild spacetime. We expect that essential qualitative features of our results in the context of entanglement harvesting will be similar to Hartle-Hawking and Unruh vacua, and we relegate explicit calculations for canonical quantization in PG coordinates for future work.

Following \cite{Aubry2014derivative,Aubry2018Vaidya,Tjoa2020vaidya}, we shall use a particular model of detector-field interaction known as  the \textit{derivative coupling} detector model. 
The reason for this choice is that the Wightman functions  \eqref{eq: wightman-boulware}-\eqref{eq: wightman-hartlehawking} have two shortcomings; they do not possess the Hadamard short-distance property \cite{Radzikowski1996,Khavkine2015}, and they have an IR ambiguity associated with massless fields in two-dimensional QFT with no boundary conditions. 
Instead, since we are interested in only the two-point functions evaluated along the support of each detector, we will only need to calculate the pullback of the two-point functions along the detectors' trajectories, and consider the proper time derivatives associated with the two trajectories $\sx(\tau)$ and $\sx'(\tau')$:
\begin{align}
    \mathcal{A}_\alpha(\sx(\tau),\sx'(\tau')) &= \Tr\rr{\pd_\tau\hat\phi(\sx(\tau))\pd_{\tau'}\hat\phi(\sx'(\tau'))\ket{0_\alpha}\!\bra{0_\alpha}}\,.
    \label{eq: derivative-twopoint-distributions}
\end{align}
The proper time derivatives remove the IR ambiguity from the Wightman function and the resulting two-point functions mimic the short-distance behavior of the Wightman distribution in (3+1) dimensions. 
It also retains all other essential features such as invariance under time translation generated by the respective timelike Killing fields that define each vacuum state. 
Past results have suggested that qualitatively similar results to the linear amplitude coupling model (without proper time derivative) are obtained in flat space and (1+1)-dimensional spacetimes with moving mirrors \cite{pozas2016entanglement, cong2019entanglement}.

More explicitly, the proper time derivative two-point function reads
\begin{subequations}
\begin{align}
    \mathcal{A}\ts{B}(\tau,\tau') 
    &= -\frac{1}{4\pi}\left[\frac{\dot{u}\dot{u}'}{(u-u'-\ii\epsilon)^2}+\frac{\dot{v}\dot{v}'}{(v-v'-\ii\epsilon)^2}\right]\,,
    \label{eq: derivative-boulware}\\
    \mathcal{A}\ts{U}(\tau,\tau') 
    &= -\frac{1}{4\pi}\left[\frac{\dot{U}\dot{U}'}{(U-U'-\ii\epsilon)^2}+\frac{\dot{v}\dot{v}'}{(v-v'-\ii\epsilon)^2}\right]\,,
    \label{eq: derivative-unruh}\\
    \mathcal{A}\ts{H}(\tau,\tau') 
    &= -\frac{1}{4\pi}\left[\frac{\dot{U}\dot{U}'}{(U-U'-\ii\epsilon)^2}+\frac{\dot{V}\dot{V}'}{(V-V'-\ii\epsilon)^2}\right] \,.
    \label{eq: derivative-HHI}
\end{align}
\end{subequations}
where we used the shorthand $\mathcal{A}_\alpha(\tau,\tau')\equiv \mathcal{A}_\alpha(\sx(\tau),\sx'(\tau'))$, $\dot{y}\equiv \pd_\tau [y(\tau)]$, and $\dot{y}'\equiv \pd_{\tau'}[y(\tau')]$. 
We stress that in general $\tau$ and $\tau'$ in the derivative coupling Wightman functions are proper times associated with \textit{distinct trajectories} $\sx(\tau)$ and $ \sx'(\tau')$; thus in general $\dd \tau/\dd\tau'\neq 1$ due to relative gravitational and kinematic redshifts of the two trajectories. For the rest of this paper we will denote  $\mathcal{A}_\alpha(\tau,\tau')\equiv \mathcal{A}_\alpha(\sx(\tau),\sx'(\tau'))$ in Eq.~\eqref{eq: derivative-boulware}-\eqref{eq: derivative-HHI} the vacuum two-point functions and not the original Wightman functions  \eqref{eq: wightman-boulware}-\eqref{eq: wightman-hartlehawking}.

\section{Setup}
\label{sec: UDW-model}

In this section, we review the Unruh-DeWitt (UDW) detector model. 
We employ the derivative coupling model so that we can avoid the IR-ambiguity, which appears in the case of a linearly coupled UDW detector in $(1+1)$-dimensional spacetimes (see \cite{Tjoa2020vaidya} and the references therein). 
To examine the properties of correlations extracted from the vacuum, we will use the perturbation theory to obtain the density operator $\rho\ts{AB}$ of two detectors.

\subsection{Derivative coupling UDW model}
Let us consider two observers, Alice and Bob, each of them carrying a pointlike UDW detector. The detector consists of a two-level quantum system with the energy gap $\Omega$, interacting locally with the quantum scalar field along the detector's trajectory. 
In this case, we are interested in the pullback of the field operator along the detector's trajectory $\hat \phi(\sx_j (\tau_j))$, where $\sx_j(\tau_j)$ denotes the trajectory of each detector parametrized by proper time $\tau_j$. 
The interaction Hamiltonian of the detector and the field is conveniently described in terms of the detector's proper time, given by
\begin{align}
    \hat H_j^{\tau_j}(\tau_j) &= \lambda_j \chi_j(\tau_j) \hat \mu_j(\tau_j) 
    \otimes \partial_{\tau_j} \hat \phi(\sx_j(\tau_j))\hspace{0.5cm} j\in \{ \text{A, B} \}\,, 
\end{align}
where the monopole moment $\hat\mu_j$ and the switching function of each detector is given by
\begin{align}
    \hat\mu_j(\tau_j) &=\ket{e_j}\!\bra{g_j}e^{\ii\Omega_j\tau_j } + \ket{g_j}\!\bra{e_j}e^{-\ii\Omega_j\tau_j }\,,\\
    \chi_j(\tau_j) &= \exp\rr{-\frac{(\tau_j-\tau_{j,0})^2}{\sigma_j^2}}\,,
\end{align}
where $\ket{g_j}$ and $\ket{e_j}$ are ground and excited states of detector $j$. 
In what follows we will assume that the detectors are identical, in that it has the same coupling strength $\lambda_j=\lambda$, energy gap $\Omega_j=\Omega$ and switching duration $\sigma_j=\sigma$ in their own proper frames. The parameter $\tau_{0,j}$ denotes the peak of the switching function. 
Note that this Hamiltonian generates time translation with respect to $\tau_j$.

The total interaction Hamiltonian of the system is conveniently written in terms of a common coordinate time $t$ (in this work we will consider either $t=t\ts{s},t\ts{PG}$) as
\begin{align}
    \hat H\ts{I}^t(t)
    &= 
    \dfrac{\dd \tau\ts{A}}{\dd t}
    \hat H\ts{A}^{\tau\ts{A}}(\tau\ts{A}(t)) 
    +
    \dfrac{\dd \tau\ts{B}}{\dd t}
    \hat H\ts{B}^{\tau\ts{B}}(\tau\ts{B}(t)) .
\end{align}
where we have used the time-reparametrization property \cite{Pablo2018rqo, Tales2020GRQO}. 
The time evolution is given by the unitary
\begin{align}
    \hat U\ts{I}&=
    \mathcal{T} 
    \exp
    \kagikako{
        -\ii \int_{-\infty}^\infty \!\!\dd t\,\hat H^t\ts{I} (t)
    } \label{eq:time-evolution operator}\,,
\end{align}
where $\mathcal{T}$ is a time-ordering symbol. 
For sufficiently weak coupling the time evolution operator can be expanded as a Dyson series
\begin{align}
    \hat U\ts{I}=
    \mathds{1} + \hat U^{(1)} + \hat U^{(2)} + O(\lambda^3)\,,
\end{align}
where $\hat U^{(k)}$ is of order $\lambda^k$ given by
\begin{align}
    \hat U^{(1)}&=
    -\ii \int_{-\infty}^\infty \!\!\dd t\,\hat H^t\ts{I}(t), \\
    \hat U^{(2)}&=
    - \int_{-\infty}^\infty \!\!\dd t \int_{-\infty}^t \!\!\dd t'\,\hat H^t\ts{I}(t) \hat H^{t'}\ts{I}(t')\,.
\end{align}

Our primary interest in the entanglement harvesting protocol is to extract entanglement from a vacuum using spacelike separated detectors when they are initially uncorrelated. To this end we set the initial density operator $\rho_0$ of the total system to be the product state
\begin{align}
    \rho_0=
    \ket{ g\ts{A} } \bra{ g\ts{A} }
    \otimes \ket{ g\ts{B} } \bra{ g\ts{B} }
    \otimes \ket{0_\alpha } \bra{0_\alpha}\,,\hspace{0.5cm} \alpha \in \{ \text{B, U, H} \}.
\end{align}
Then the total density operator $\rho\ts{tot}$ after the time-evolution $\hat U\ts{I}$ is 
\begin{align}
    \rho\ts{tot}
    = \hat U\ts{I}^{\phantom{\dagger}} \rho_0 \hat U\ts{I}^\dag 
    &= \rho_0 + \sum_{i+j=1}^2 \rho^{(i,j)}
    + O(\lambda^3)\,,
\end{align}
where we defined $\rho^{(i,j)}\coloneqq \hat U^{(i)} \rho_0 \hat U^{(j)\dagger} $.
The composite density operator for two detectors, $\rho\ts{AB}$ is obtained by tracing out the field: $\rho\ts{AB}=\Tr_\phi [\rho\ts{tot}]$. Note that after tracing out the degrees of freedom, $ \rho^{(1,0)} $ and $ \rho^{(0,1)}$ do not contribute to the density matrix $\hat\rho\ts{AB}$ due to vanishing one-point vacuum correlation functions.

By choosing bases $\ket{g\ts{A}} \ket{g\ts{B}}=[1,0,0,0]^\top$, $\ket{g\ts{A}} \ket{e\ts{B}}=[0,1,0,0]^\top$, $\ket{e\ts{A}} \ket{g\ts{B}}=[0,0,1,0]^\top$, and $\ket{e\ts{A}} \ket{e\ts{B}}=[0,0,0,1]^\top$, $\rho\ts{AB}$ takes the following form:
\begin{align}
    \rho\ts{AB}
    &= \left[
    \begin{array}{cccc}
    1-\mathcal{L}\ts{AA} - \mathcal{L}\ts{BB} &0 &0 & \mathcal{M}^* \\
    0 &\mathcal{L}\ts{BB} &\mathcal{L}\ts{BA} &0  \\
    0 &\mathcal{L}\ts{AB} &\mathcal{L}\ts{AA} &0  \\
    \mathcal{M} &0 &0 & 0
    \end{array}
    \right]
    +O(\lambda^4)\,, \label{eq:density matrix}
\end{align}
The matrix elements are given by
\begin{align}
    &\mathcal{L}_{ij}
    =\lambda^2 
    \int_{-\infty}^\infty \dd \tau_i 
    \int_{-\infty}^\infty \dd \tau_j'\,
    \chi_i(\tau_i) 
    \chi_j(\tau_j')
    e^{ -\ii\Omega (\tau_i - \tau_j' ) }\notag\\
    &\hspace{3cm}\times\mathcal{A}_\alpha(\sx_i(\tau_i) ,\sx_j(\tau_j'))\,, \label{eq:Lij}\\
    &\mathcal{M}=
    -\lambda^2 
    \int_{-\infty}^\infty \dd\tau\ts{A}
    \int_{-\infty}^\infty \dd\tau\ts{B}\,
    \chi\ts A(\tau\ts{A}) 
    \chi\ts B(\tau\ts{B})
    e^{ \ii\Omega (\tau\ts{A} + \tau\ts{B}) } \notag \\
    &\hspace{10mm}\times 
    \bigg[
        \Theta \big( t(\tau\ts{A}) - t(\tau\ts{B}) \big)
        \mathcal{A}_\alpha \big(\sx\ts{A}(\tau\ts{A}), \sx\ts{B}(\tau\ts{B}) \big)\notag\\
        &\hspace{1.2cm}+\Theta \big( t(\tau\ts{B}) - t(\tau\ts{A}) \big)
        \mathcal{A}_\alpha \big( \sx\ts{B}(\tau\ts{B}) , \sx\ts{A}(\tau\ts{A}) \big)
    \bigg]\,,
    \label{eq:nonlocal M}
\end{align}
where $\Theta(z)$ is Heaviside step function and the pullback of the Wightman function along the trajectories of both detectors reads
\begin{align}
    \mathcal{A}_\alpha \big(\sx_i(\tau_i) ,\sx'_j(\tau'_j) \big) = \bra{0_\alpha}{\partial_{\tau_i} \hat \phi(\sx_i(\tau_i)) \partial_{\tau'_j} \hat \phi(\sx_j(\tau'_j))}\ket{0_\alpha}\,.
\end{align}
$\mathcal{L}\ts{AA}$ and $\mathcal{L}\ts{BB}$ are the transition probabilities of Alice and Bob, respectively. $\mathcal{M}$ and $\mathcal{L}\ts{AB}(=\mathcal{L}\ts{BA}^*)$ corresponds to the nonlocal terms that simultaneously depend on both trajectories; $\mathcal{M}$ is responsible for entangling two detectors and $\mathcal{L}\ts{AB}$ is used for calculating the mutual information. 

Let us comment on the choice of coordinate systems. The coordinate system $x^\mu$ is chosen in such a way that it specifies the coordinates of two detectors. 
In $(1+1)$-dimensional Schwarzschild spacetime, such a coordinate system could be the Schwarzschild, Eddington-Finkelstein, Kruskal-Szekeres, PG coordinate systems, etc. 
For the two static detectors case considered in \cite{Tjoa2020vaidya}, any one of the coordinate systems above can be used. However, this is not true when one of the detectors is free-falling and enters the black hole; Schwarzschild coordinates cannot be used since it prevents us from analyzing the horizon-crossing moment. For this purpose, the coordinate system adapted to free-falling observers will be the simplest for our purposes both conceptually and numerically.  The calculations of the geodesic equation for the free-falling trajectories in terms of the double null coordinates are given in Appendix~\ref{appendix: geodesic}.

In this paper, we will employ the PG coordinate system to examine the harvesting protocol, that is, the time parameter $t$ used in \eqref{eq:time-evolution operator} will be $t\ts{PG}$, and so the Heaviside step functions in \eqref{eq:nonlocal M} become 
\begin{align}
    \Theta(t\ts{PG}(\tau\ts{A})-t\ts{PG}(\tau\ts{B})),~
    \Theta \big(t\ts{PG}(\tau\ts{B}) - t\ts{PG}(\tau\ts{A}) \big)\,.
\end{align}
This choice is possible since the time ordering is preserved when the detectors have negligible spatial extent \cite{EMM2021brokencovariance}.

Let us now move on to the measures of correlation, concurrence \cite{Wotters1998entanglementmeasure} and mutual information \cite{nielsen2000quantum}.  We use concurrence as a measure of the amount of entanglement extracted from a vacuum. 
Given the density matrix \eqref{eq:density matrix}, concurrence $\mathcal{C}[\rho\ts{AB}]$ takes the form \cite{Horodecki996separable,smith2016topology}
\begin{align}
    \mathcal{C} [\rho\ts{AB}] 
    = 2\max\{0,\, |\mathcal{M}| - \sqrt{\mathcal{L}\ts{AA} \mathcal{L}\ts{BB} }\}+O(\lambda^4).
\end{align}
Although there are other entanglement measures such as negativity \cite{Vidal2002negativity, pozas2015harvesting}, concurrence gives a nice intuition for entanglement harvesting; entanglement can be extracted when the nonlocal $|\mathcal{M}|$ is greater than the local noise contribution $\sqrt{ \mathcal{L}\ts{AA} \mathcal{L}\ts{BB} }$. 
In this sense $\mathcal{M}$ is responsible for entanglement extraction and the probabilities $\mathcal{L}\ts{AA}, \mathcal{L}\ts{BB}$ act as a noise. 

We are also interested in mutual information $I[\rho\ts{AB}]$, which tells us how much general correlations, including classical ones, are extracted. It is defined by \cite{nielsen2000quantum}
\begin{align}
    I[\rho\ts{AB} ]
    \coloneqq S[\rho\ts{A} ]+S[\rho\ts{B} ]- S[\rho\ts{AB} ],
\end{align}
where $S[ \rho ]\coloneqq-\Tr[ \rho \ln \rho ]$ is the von Neumann entropy. 
In the case of \eqref{eq:density matrix}, the mutual information is known to be \cite{pozas2015harvesting}
\begin{align}
    I[\rho\ts{AB} ] 
    &= \mathcal{L}_+\ln \mathcal{L}_+ 
    + \mathcal{L}_- \ln \mathcal{L}_- \notag \\
    &- \mathcal{L}\ts{AA} \ln \mathcal{L}\ts{AA}
    - \mathcal{L}\ts{BB} \ln \mathcal{L}\ts{BB} 
    + O(\lambda^4), 
\end{align}
where 
\begin{align}
     \mathcal{L}_{\pm} 
     &= \frac{1}{2} 
     \kako{
        \mathcal{L}\ts{AA}
        + \mathcal{L}\ts{BB} 
        \pm 
        \sqrt{
            ( \mathcal{L}\ts{AA} - \mathcal{L}\ts{BB} )^2 
            + 4| \mathcal{L}\ts{AB} |^2
        }
     }.
\end{align}
If the two detectors do not have entanglement but have nonzero mutual information, then the correlations between them must be either classical correlation or nonentanglement quantum correlations such as discord \cite{Zurek2001discord,Henderson2001correlations}.


\subsection{Causality for the two detectors}

In \cite{pozas2015harvesting,pozas2016entanglement}, there is an emphasis on the fact that entanglement harvesting protocol is most relevant when the two detectors in question are spacelike separated (or at least approximately so). This is reasonable because generic interactions will produce correlations between detectors. This is true for the Unruh-DeWitt model for arbitrary choice of physical parameters such as energy gap and switching duration. This is because for generic interactions, communication between detectors will generate entanglement unless the channel on $\rho\ts{AB}$ induced from the global detector-field unitary is entanglement breaking \cite{Horodecki2003entanglement-break, Simidzija2018no-go}. For instance, in the case of  degenerate detectors or instantaneous switching,  entanglement extraction from the QFT vacuum is impossible \cite{pozas2017degenerate, Simidzija2018no-go}. It was also shown in \cite{Simidzija2018no-go} how entanglement extraction from the vacuum might be assisted by allowing some form of communication between the two detectors. 

The point is that for generic setups of detector-field interaction, the quantum channel $\Phi(\rho{\ts{AB}})=\Tr_\phi(\hat U\rho_0\hat U^\dagger)$ is not an entanglement-breaking channel. Consequently, in general the nonzero concurrence (and mutual information) harvested in this protocol will be a mixture of contributions purely from the field vacuum extraction and also from the communication between the two detectors. Since the setup we consider is both perturbative and generic (i.e., not in the entanglement-breaking regimes such as where detectors have degenerate gaps or the switching is instantaneous), nonzero concurrence and mutual information would suggest that the purely vacuum contribution is also nonzero. Therefore, we do not attempt to enforce that the detectors are strictly spacelike separated unless otherwise stated. 

In the case where the causal relations are important, we will provide a measure of how causally disconnected the two detectors are. Following \cite{Causality2015Eduardo,Tjoa2020vaidya}, we introduce a signalling estimator $\mathcal{E}$ which serves as a primitive tool to analyze the causal relations between the two detectors. We take the signalling estimator to be\footnote{Strictly speaking we only need the modulus $|\mathcal{E}|$ since what is more relevant is the spatial interval or region where $\mathcal{E}$ is approximately zero: the nonzero value of $\mathcal{E}$ itself is secondary.}
\begin{align}
    \label{estimator}
    \mathcal{E} &\coloneqq
    \frac{\lambda^2}{2} \text{Im}
    \bigg(
        \int_{-\infty}^\infty \dd \tau\ts{A} 
        \int_{-\infty}^\infty \dd \tau\ts{B}\,
        \chi(\tau\ts{A})
        \chi(\tau\ts{B})\notag\\
        &\hspace{1cm}\times \bra{0}
        [ \partial_{\tau\ts{A}} \hat \phi (\sx\ts{A}( \tau\ts{A} )) ,\partial_{\tau\ts{B}} \hat \phi (\sx\ts{B}( \tau\ts{B} )) ] \ket{0}
    \bigg)\,.
\end{align}
We have removed the subscript $\alpha$ from the vacuum state since the field commutators are $c$-numbers. This estimator is useful for the following reason: due to finite switching times for both detectors, it is classically challenging to determine if Alice is in the causal complement of Bob or not since it will require nontrivial ray tracing for the entire spatiotemporal support of both detectors, even if the detectors are pointlike. This provides us with a relatively cheap measure of how spacelike/timelike two detectors are; the main drawback is that it does not allow us to clearly quantify how much of the correlations harvested is  due to communication-assisted contributions and how much is the truly vacuum harvesting part. 

{A cautionary note is in order here.  The signalling estimator is crude insofar as it is not quite an entanglement monotone; this will become apparent in what follows.  Thus nonzero $|\mathcal{E}|$ can only indicate that some of the concurrence is due to field-mediated communication channel between Alice and Bob. Conversely, zero   $|\mathcal{E}|$ guarantees that the entanglement harvested is purely from the vacuum of the quantum field as the detectors are causally disconnected. Likewise, small $|\mathcal{E}|$ is indicative that most of the harvested entanglement is not due to a communication channel.
The sign of $\mathcal{E}$ is secondary\footnote{Indeed, the absolute value $|\mathcal{E}|$ is the definition of signalling estimator in \cite{Causality2015Eduardo}.} and in what follows we shall plot $|\mathcal{E}|$ whenever it is conceptually clearer to do so.}

Note that the signalling estimator $\mathcal{E}$ is in terms of the commutator of the proper time derivative of the field along the two detectors' trajectories instead of the field commutator  $[\hat\phi(\sx\ts{A}(\tau\ts{A})), \hat\phi(\sx\ts{B}(\tau\ts{B}))]$. This is because for the derivative-coupling Unruh-DeWitt model, one can show that that the leading order correction\footnote{Here $\rho^{(2)}\ts{AB}$ contains all terms of order $\lambda^2$ in Eq.~\eqref{eq:density matrix}.} to Bob's density matrix  $\rho_{\ts{B}}^{(2)}\coloneqq \Tr_{\ts A}\rho_{\ts{AB}}^{(2)}$ due to Alice's detector depends only on the commutator 
$[ \partial_{\tau\ts{A}} \hat \phi (\sx\ts{A}( \tau\ts{A} )) ,\partial_{\tau\ts{B}} \hat \phi (\sx\ts{B}( \tau\ts{B} )) ]$. 
The calculation proceeds analogously up to Eq.~(24) in \cite{Causality2015Eduardo}. Furthermore, since the signalling component depends on proper time derivatives, it means that the communication contribution of the harvesting protocol mimics (3+1)-dimensional setting, in that the commutator only has support on the light cone. Therefore, communication mediated by the massless scalar field will only occur if the supports of the switching functions of Bob's detector overlap with the lightlike boundary of Alice's causal past/future and vice versa. This means that they only can can communicate via the scalar field if the support of the Gaussian of one detector intersects the causal past/future of the other detector. 

Finally, we remark that the Gaussian switching can be effectively taken to have compact support despite the infinite exponential tails. {We  define   \textit{strong support}   to be the interval $[-5\sigma+\tau_{j,0},5\sigma+\tau_{j,0}]$, where $\tau_{j,0}$ are the Gaussian peaks defined in each detector's rest frame. The switching function $\chi(\tau_j)$ can be taken to be negligible outside of this interval.
} This sort of strong support approximation has been shown to be reasonable for detector-field interaction studies \cite{Causality2015Eduardo, Tjoa2020vaidya}.

\section{Results and discussions}
\label{sec: results}

\begin{figure*}[t]
    \centering
    \includegraphics[width=16cm]{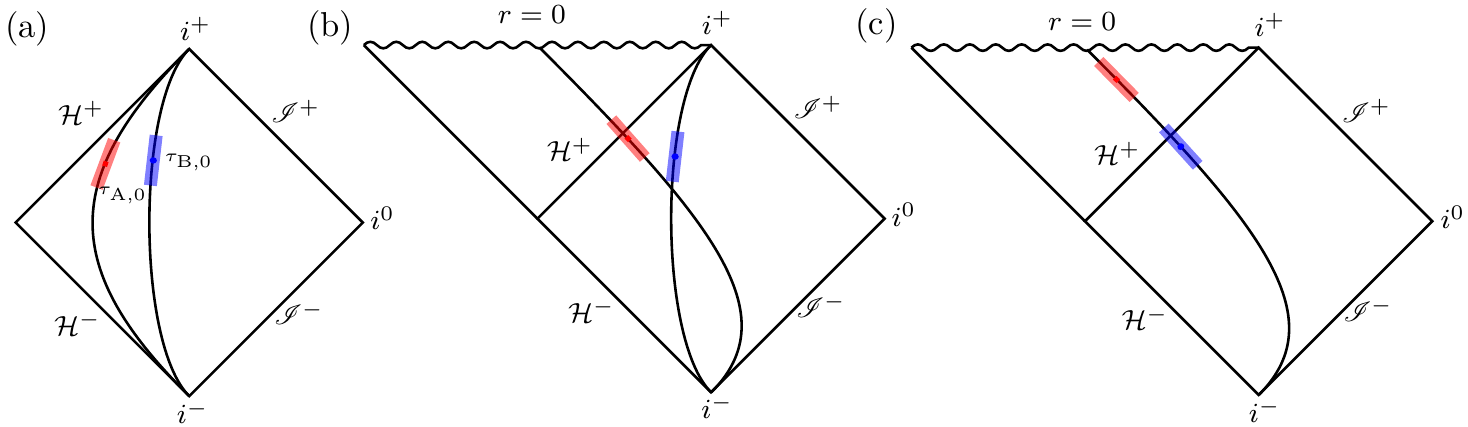}\\
    \caption{~Penrose diagrams for three cases: (a) SS, (b) FS, and (c) FF.}
\label{fig:Penrose diagram for three cases}
\end{figure*}


In this section we present our main results for entanglement and mutual correlation harvesting for various parameter choices and detector trajectories. We define $d(r_i,r_j)$ to be the proper distance between the coordinate radii $r_i,r_j$, and   write $d\ts{AB} \coloneqq d(r\ts A,r\ts B)$ for the proper separation between two detectors.  

It is clear that for two static detectors $d\ts{AB}$ is unambiguous, {but for free-falling detectors the proper distance between them  changes with time.} Therefore, there is a need to find some sort of effective proper distance that works for free-falling scenarios. For this purpose, we use the locations of the peaks of Alice and Bob's Gaussian switching functions (given in terms of the peak proper times $\tau\ts{A,0}$ and $\tau\ts{B,0}$ of the detectors) as reference points as follows.
\begin{enumerate}[leftmargin=*,label=(\alph*)]
    \item If Alice and Bob are static (we will call this the \textbf{SS scenario}), we take $d\ts{AB}$ to be measured when both peaks are at some constant-$t\ts{PG}$ slice. Since both are static, their separation $d\ts{AB}$ computed this way is also valid even if the switching peaks are translated along their respective trajectories.

    \item If Alice is free-falling and Bob is static (we will call this the \textbf{FS scenario}), we take $d\ts{AB}$ to be measured with respect to the peaks of their Gaussian switching, thus effectively locating Alice at $r\ts{A} = r(\tau\ts{A,0})$ and Bob at some fixed $r\ts{B}$. We stress that both detector trajectories are parametrized by different proper times due to gravitational redshift, i.e., $\dd\tau\ts{A}/\dd\tau\ts{B}\neq 1$.
    
    \item If both detectors are free-falling from infinity (we will call this the \textbf{FF scenario}) initially at rest, then the proper distance is given by $d\ts{AB}=|r(\tau\ts{A,0})-r (\tau\ts{A,0}+\text{const.})|$ with the constant to be determined. This is because in this case both detector trajectories can be parametrized by the same proper time and hence the proper distance is completely controlled by the difference $\tau\ts{B,0}-\tau\ts{A,0}=\text{constant}$.
\end{enumerate}
This means that the role of the free-falling motion in contrast to the static one is encoded in the support and peak of the switching functions. The three scenarios are depicted in Fig.~\ref{fig:Penrose diagram for three cases}.

{We will compute all quantities}  in units of the switching width $\sigma$. Since the coupling strength for the derivative UDW model has units of  $[\text{Length}]^{(n-1)/2}$ where $n$ is the number of spatial dimensions,   we can define $\tilde\lambda\coloneqq \lambda\sigma^{(n-1)/2}$. In (1+1) dimensions, this gives $\tilde\lambda=\lambda$ (i.e. $\lambda$ is already dimensionless) but we will use $\tilde\lambda$ in this section to remind ourselves that the coupling strength of the (derivative) UDW model is dimension dependent. 

All the results below are obtained numerically using the technique involving numerical contour integration outlined in \cite{Tjoa2020vaidya}, with a modification for the evaluation of the nonlocal term $\mathcal{M}$ in \eqref{eq:density matrix}: the free-falling trajectory introduces some numerical instability that makes it difficult to work with the  Heaviside step function directly. Consequently, the computation is done by approximating the step function using a smooth analytic function: for our purposes we use the fact that
\begin{align}
    \Theta(z)=\lim_{k\to\infty}\rr{\dfrac{1}{2}+\dfrac{1}{2}\tanh kz}\,,
\end{align}
and define an \textit{approximate step function}\footnote{The analysis of this technique is given in \cite{Tjoa2021mathematica}, which includes the performance of this approximation along with other possible choices of analytic functions and variations of the contour. We also note that the numerical evaluation is done using \textit{Mathematica} 10 \cite{Mathematica}, as it is (surprisingly) more stable than the newer versions and in some cases the newer versions may even compute the wrong answers on physical grounds.} to be 
\begin{align}
    \Theta_k(z)\coloneqq \frac{1}{2}+\frac{1}{2}\tanh kz
\end{align}
where $k$ is fixed but sufficiently large. The choice of $k$ will in fact be dependent on the choice of the contour size (i.e. the value of $\epsilon$ in the $\ii \epsilon$ prescription), which for generic situations requires small $\epsilon$ for large $k$.

\subsection{Free-falling Alice, static Bob (FS)}

\begin{figure*}[tp]
    \centering
    \includegraphics[width=\textwidth]{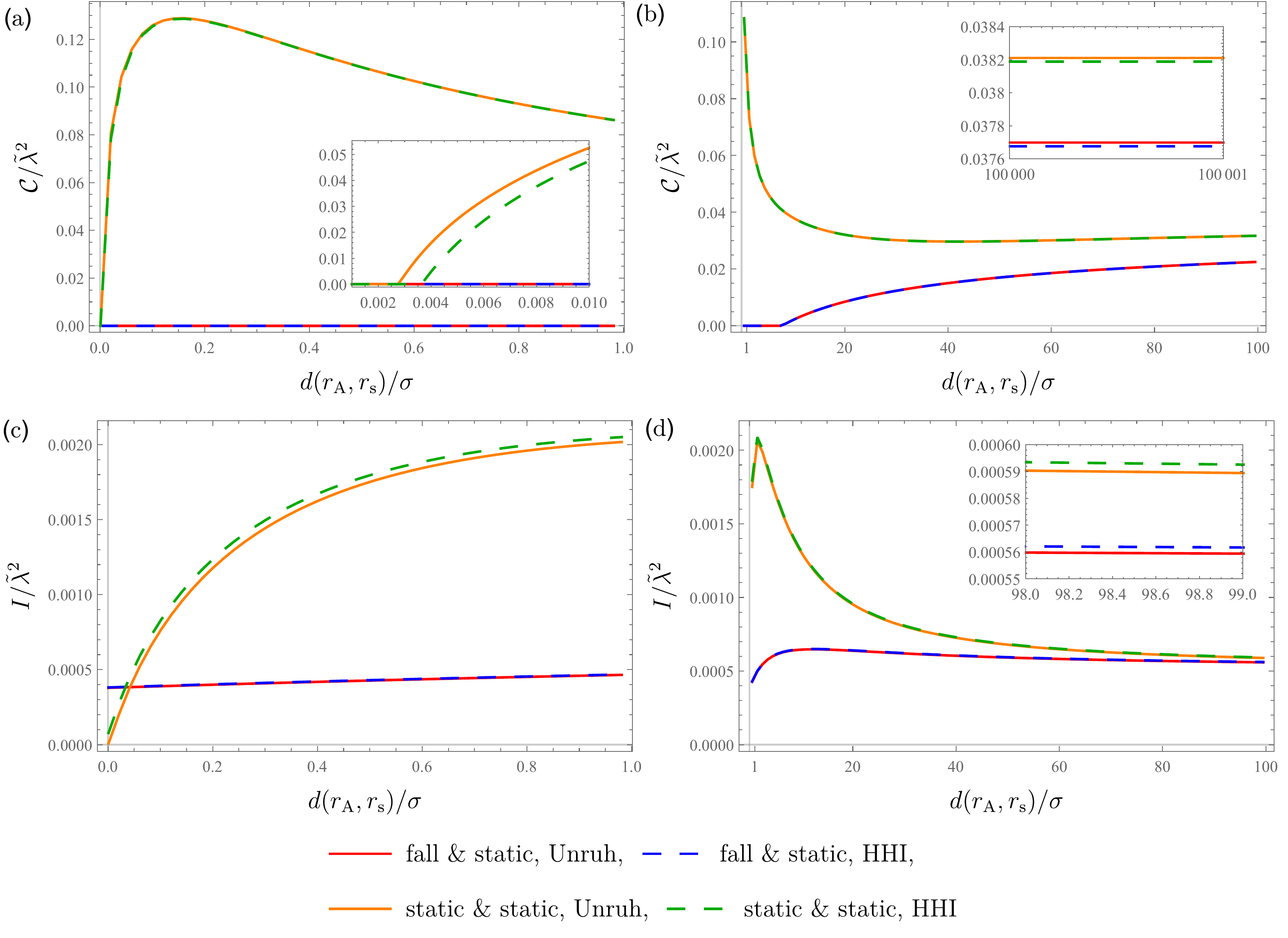}\\
    \caption{
    ~ Concurrence and mutual information are plotted as a function of  the effective proper distance of Alice from the horizon for both the SS and FS scenarios. 
    Here we chose $\Omega \sigma=2, M/\sigma=5, d\ts{AB}/\sigma=2$.
    (a) Concurrence near the horizon, where $d(r\ts{A}, r\ts{s})/\sigma \in [0.001, 1]$. 
    (b) Concurrence further away from the horizon at $d(r\ts{A}, r\ts{s})/\sigma \in [1, 100]$ with the inset covering very far regime $d(r\ts{A}, r\ts{s})/\sigma \in [10^5, 10^5+1]$.
    (c) Mutual information near the horizon, where $d(r\ts{A}, r\ts{s})/\sigma \in [0.001, 1]$. (d) Mutual information further away from the horizon at $d(r\ts{A}, r\ts{s})/\sigma \in [1, 100]$.}
    \label{fig:Concurrence FSvsSS}
\end{figure*}

In Fig.~\ref{fig:Concurrence FSvsSS} we plot the concurrence and mutual information as functions of Alice's proper distance from the horizon at the exterior of the black hole for both Unruh and Hartle-Hawking vacuum states and we compare the FS and SS scenarios. 

We observe that the static-static (SS) scenario has larger concurrence than the free-falling-static (FS) scenario, and this is true even when both detectors are far from the black hole. 
The entanglement shadow near the horizon is wider for the FS case, whereas for the  SS case the shadow is much smaller in comparison [see inset in Fig.~\ref{fig:Concurrence FSvsSS}(a)].
The generic result here is that when one detector is free-falling, the bipartite entanglement harvesting is less potent than the static case. 
We will revisit this issue later in order to see to what extent this can be explained by relative velocities between the two detectors. 
We remark that the results for the SS case differ somewhat  from those a previous study of harvesting in $(1+1)$-dimensional collapsing shell spacetime
\cite{Tjoa2020vaidya} because the protocols are implemented slightly differently;  
here the switching peaks of the detectors are turned on at the same constant $t\ts{PG}$ slices, while in \cite{Tjoa2020vaidya} the detectors are turned on at the same constant proper time $\tau_0$ (in their own frames). The relative redshift factor is markedly different, and hence the entanglement shadow size is different [very small in Fig.~\ref{fig:Concurrence FSvsSS}(a) inset].

However for mutual information harvesting, notably the FS scenario can outperform the SS case {very near to} the horizon, as we show in Fig.~\ref{fig:Concurrence FSvsSS}(c). The overall behavior for mutual information harvesting is similar to concurrence in that the FS case is less efficient in extracting correlations from the field, although there is nonzero mutual information in general. This behavior is only possible because generically there is no entanglement shadow for mutual information in this framework, unlike the situation when one computes entanglement monotones such as concurrence and negativity.

\begin{figure*}[tp]
    \centering
    \includegraphics[width=\textwidth]{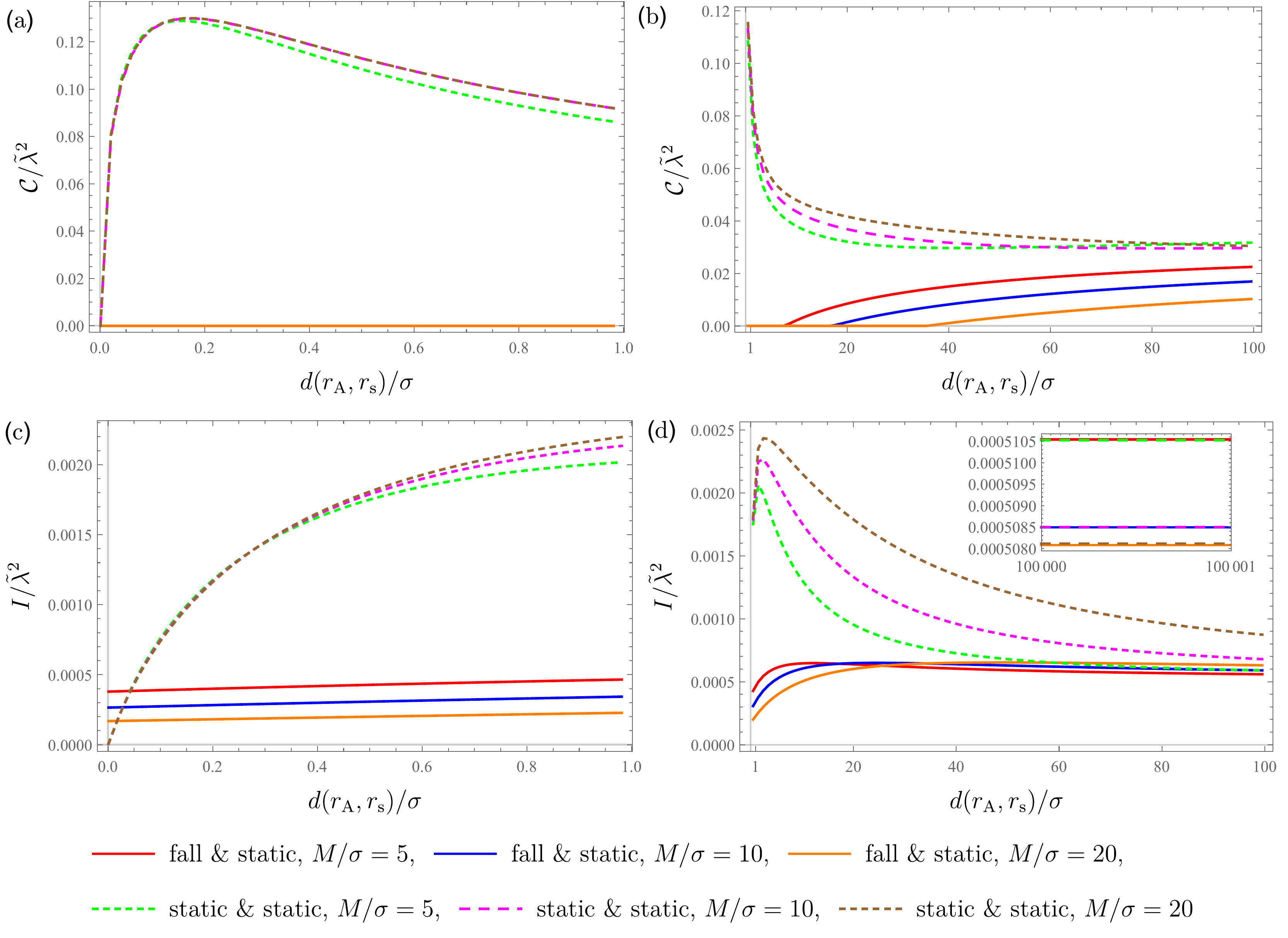}%
    \caption{Concurrence and mutual information are plotted for various black hole masses in both the FS and SS scenarios as a function of Alice's proper distance from the horizon. As a benchmark we use the  Unruh vacuum and we consider three different masses $M/\sigma=5,10,20$. We fix the other parameters as $\Omega \sigma=2, d\ts{AB}/\sigma=2$. (a) Concurrence near the horizon at $d(r\ts{A}, r\ts{s})/\sigma \in [0.001, 1]$. We see that the FS case vanishes close to the horizon. (b) Concurrence further away at $d(r\ts{A}, r\ts{s})/\sigma \in [1, 100]$. The entanglement shadow in the SS case increases with increasing $M$ \cite{Tjoa2020vaidya}. (c) Mutual information near the horizon at $d(r\ts{A}, r\ts{s})/\sigma \in [0.001, 1]$. (d) Mutual information further away at $d(r\ts{A}, r\ts{s})/\sigma \in [1, 100]$.}
    \label{fig:Concurrence different mass}
\end{figure*}

Next, we consider how the harvesting protocol depends on the black hole masses, as we show in Fig.~\ref{fig:Concurrence different mass}. In general, we see that the smaller mass black holes allow for better harvesting efficiency for both concurrence and mutual information. However, due to nontrivial roles of curvature and communication between two detectors, the variation of concurrence and mutual information as we vary detector distances from the horizon is generally not monotonic. This is especially so for mutual information, where we see that sufficiently far from the horizon, the behavior flips and detectors harvest mutual information less for smaller masses; we verified at large distances [see inset of Fig.~\ref{fig:Concurrence different mass}(d)] that the curves flip again, and we again obtain the result that larger mass leads to less correlation harvested.

A natural question that arises is to what extent the results obtained thus far depend only on the kinematic properties of the detectors (i.e., their velocities) and how much of it comes from the intrinsic properties of the background spacetime (i.e., the curvature). As it turns out, the UDW formalism is not very sensitive to spacetime curvature and much of the results here can be simulated using the corresponding flat space result, at least in the exterior geometry of the black hole sufficiently distant from the horizon. In order to better understand this kinematical aspect of the harvesting protocol, we will consider concurrence and mutual information for the Boulware vacuum and compare the results with the Minkowski vacuum analog\footnote{A similar comparison has been carried out between Rindler and Schwarzschild spacetimes \cite{Ahmadzadegan:2013iua}.}. 

For this comparison to work, we will use the notion of intrinsic relative velocities in general relativity. The idea is that since in curved spacetimes $\mathscr{M}$ the connection is not flat, one cannot compare vectors belonging to tangent spaces {at different} points directly. Consequently, in the presence of curvature one cannot na\"ively compute relative velocities between two observers at two different events $p,q$ because a vector $u^a$ in $T_p\mathscr{M}$ is not \textit{a priori} related to vectors in $T_q\mathscr{M}$. However, for spacetimes with well-defined spacelike foliations one can generalize the notion of relative velocities by making use of generalized version of spacelike simultaneity in flat space. Formally, given four-velocity $u^a\in T_p\mathscr{M}$,   spacelike simultaneity is given by the so-called \textit{Landau submanifold} $L_{p,u}$, defined via the submersion $\Phi:\mathscr{M}\to \mathbb{R}$ with $\Phi(q) = g(\exp_p^{-1}q, u) = g_{\mu\nu}(\exp_p^{-1}q)^\mu u^\nu$ where $\exp$ is the exponential map \cite{bolos2007intrinsic}. In other words,   spacelike simultaneity is defined by the spacelike hypersurface obtained as the regular level set $L_{p,u}=\Phi^{-1}(0)$. This defines the so-called \textit{kinematic relative velocity} (see \cite{bolos2007intrinsic} and references therein for other definitions of relative velocities that can be defined in general relativity).

\begin{figure*}[t]
    \centering
    \includegraphics[width=\textwidth]{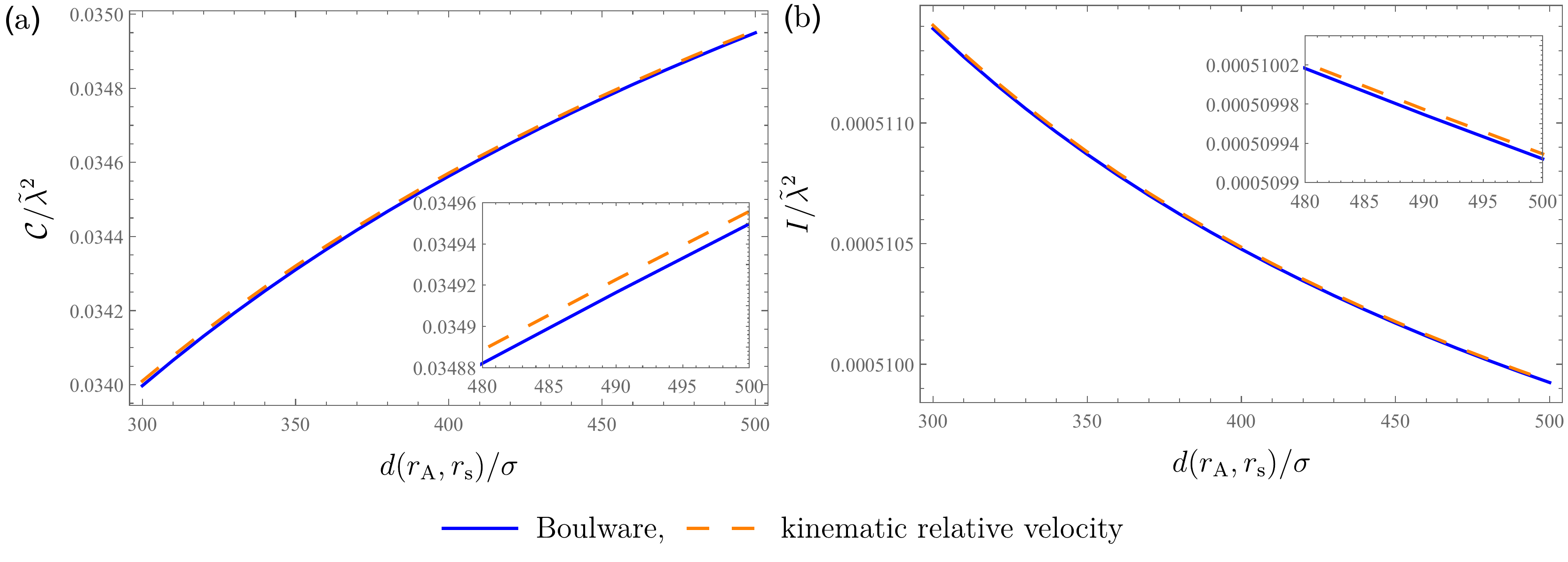}%
    \caption{~  Concurrence and mutual information for the FS scenario are plotted for the Boulware vacuum and compared to the KRV  in the Minkowski vacuum. $\Omega \sigma=2, M/\sigma=1, d\ts{AB}/\sigma=2$, and stationary and boosted inertial detectors in $(1+1)$-dimensional Minkowski spacetime with $\Omega \sigma=2, d\ts{AB}/\sigma=2$. }
    \label{fig:relative velocity}
\end{figure*}

For Schwarzschild geometry, the notion of kinematic relative velocity (KRV) can be defined using the Landau submanifold above, and  boils down to a simple formula in terms of the metric function $f(r)$. If Alice and Bob are both static observers at fixed radii, the KRV is zero. For the FS scenario where Alice is free-falling from infinity, her KRV relative to Bob is given by \cite{bolos2007intrinsic}
\begin{align}
    V\ts{A,\text{kin}}\bigr|\ts{B} &= \left.-f(r\ts{B})\sqrt{1-\frac{f(r\ts{A})^2}{E^2}}\frac{\partial}{\partial r}\right|\ts{B}\,,
\end{align}
where $E = \sqrt{1-2M/r_0}$ and $r_0$ is the initial radius of the free-falling trajectory at rest \cite{chandrasekhar1998mathematical}. In our FS scenario, we have $r_0\to \infty$ so that $E=1$ and the magnitude of the KRV is given simply $V\ts{AB} \coloneqq |V\ts{A,\text{kin}}| = f(r\ts{B})\sqrt{1- f(r\ts{A})^2}$. Note that $V\ts{AB}$ depends on $\tau\ts{A}$ since a free-falling detector has nonzero proper acceleration; in this case, Alice's proper acceleration can be shown to be [see Eq.~\eqref{eq:acceleration} in Appendix~\ref{appendix: geodesic}]
\begin{align}
    a(\tau\ts{A}) \coloneqq  \sqrt{a_\mu a^\mu} = -\dfrac{M}{r(\tau\ts{A})^2}\,.
\end{align}
If the detectors are far enough from the black hole and/or the support of the Gaussian is sufficiently small, then the variation of Alice's radial velocity across the Gaussian support can be considered approximately constant, equal to the value at Alice's Gaussian peak. In this case, the KRV is given approximately by $V\ts{AB,0} \coloneqq f(r\ts{B})\sqrt{1- f(r(\tau\ts{A,0}))^2}$ where $\tau\ts{A,0}$ is the Gaussian peak of Alice. We can then compare the concurrence and mutual information of the corresponding scenario in {Minkowski space} where Alice has relative velocity $V\ts{AB}$ with respect to Bob for the same derivative coupling UDW model.\footnote{All we need to change is the definition of the null coordinates into $u,v = t\mp r$ in the definition of Wightman distribution for the Boulware vacuum \eqref{eq: derivative-boulware}.}

We compare the concurrence and mutual information in the FS scenario in the Boulware vacuum against the corresponding Minkowski vacuum scenario with the same constant relative velocity $V\ts{AB,0}$ in Fig.~\ref{fig:relative velocity}. The flat space version (shown in orange in Fig.~\ref{fig:relative velocity}) corresponds to Bob at rest in an inertial frame and Alice boosted away from Bob, with $d\ts{AB}=2\sigma$ measured from the peaks of both Gaussian switching functions. The relevant scale is given by $a(\tau\ts{A,0}) \sigma \approx 10^{-5}\ll 1$ (thus the constant velocity approximation is valid across the Gaussian support) and $V\ts{AB,0}\lesssim 0.08$ which is almost in the relativistic regime.  Observe that for this setup, most of the correlations at $d(r\ts{A},r\ts{s})\sim 300\sigma-500\sigma$ can be accounted for by the correct relative velocity alone (hence purely kinematic). Therefore, the relative motion between the detectors (measured by KRV) is the most relevant physics that explains why   correlations in the FS scenario are consistently lower than the SS counterpart in Figs.~\ref{fig:Concurrence FSvsSS} and~\ref{fig:Concurrence different mass}. At distances much farther than $500\sigma$, the concurrence and mutual information harvesting are practically indistinguishable from flat space. As we approach the horizon the correlation harvested will start to be different for fixed $\sigma$ as more non-uniformity in the accelerated motion is captured by the Gaussian support.

We remark that this does not mean the effect of gravitational field is absent from the harvesting protocol; in fact our analysis based on the kinematic relative velocity is a manifestation of local flatness and the equivalence principle, since  small enough Gaussian support is equivalent to looking at a small enough region of spacetime (the detector is already pointlike). The fact that the black hole is present will be manifest in other ways: for instance, insofar as Alice cannot signal to detectors near future null infinity $\mathscr{I}^+$ once Alice falls into the black hole, or that Alice will hit the singularity in finite proper time. Also, by putting Bob in static trajectory such that the switching peak is sufficiently near to future timelike infinity $i^+$ while Alice is inside the black hole, it is {guaranteed} that any correlation harvested is from the QFT vacuum (without a  communication component) since they are both causally disconnected by the future horizon $\mathcal{H}^+$. In \cite{Tjoa2020vaidya} it was shown how signalling estimator for static detectors is already generally nontrivial in some finite region in the exterior: since the field commutator is state-independent, the nontrivial signalling is gravitational in nature as the classical solutions to the Klein-Gordon equation {depend on} curvature (reflected by the nontrivial wave operator $\nabla_\mu\nabla^\mu$).

\begin{figure*}[t]
\begin{center}

    \begin{tabular}{c}
    \begin{minipage}[c]{0.38\linewidth}
    \begin{center}
        \includegraphics[clip,width=5.5cm]{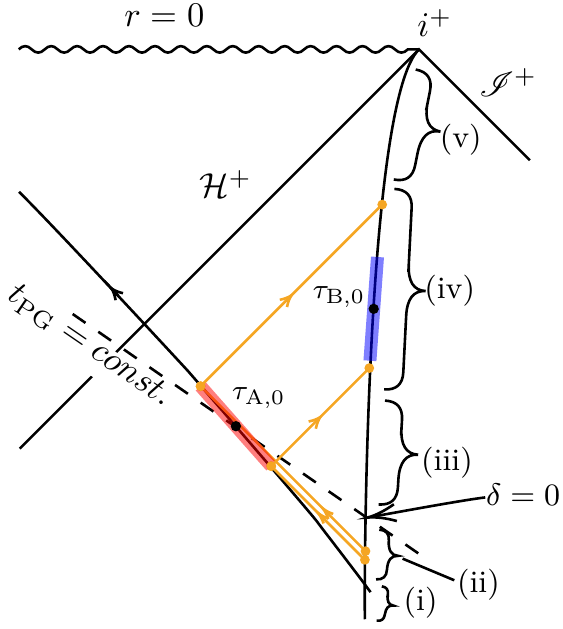}
        \hspace{1.6cm}(a)
    \end{center}
    \end{minipage}

    \begin{minipage}[c]{0.55\linewidth}
    \begin{center}
        \includegraphics[clip,width=\textwidth]{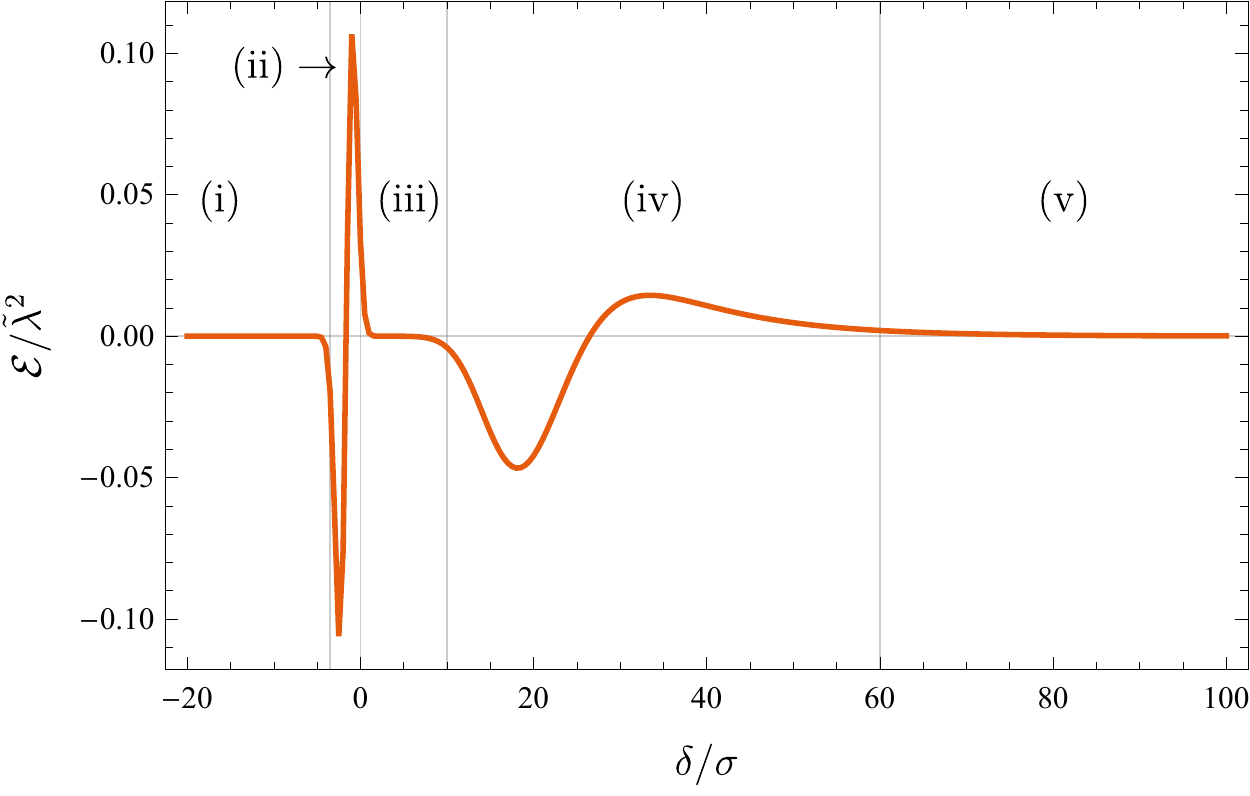}
        \hspace{1.6cm}(b)
    \end{center}
    \end{minipage}

\end{tabular}
\caption{~Penrose diagram of the schematics of Alice and Bob's positions,  and the corresponding signalling estimator between Alice and Bob as a function of the ``time-delay'' parameter for the FS scenario. Here we chose $M/\sigma=5, d\ts{AB}/\sigma=5$ and $d(r\ts{A},r\ts{s})/\sigma=1$. The red and blue stripes denote the strong support of Alice's and Bob's Gaussian switching, and the orange lines denote the light rays that emanate or arrive at the endpoints of Alice's strong support.}
\label{fig:FS signalling}
\end{center}

\end{figure*}

\subsection{Dependence of harvesting with signalling between detectors}

Our analysis so far has been focused on the detector trajectories, regardless of whether the detectors harvest correlations purely from the vacuum or potentially assisted by communication. We would like to understand to what extent the harvesting protocol in this particular setup is assisted by a communication channel between the detectors mediated by the quantum field. 

To this end, we consider the signalling estimator for Alice and Bob's trajectories, as shown in Fig.~\ref{fig:FS signalling}. The orange lines denote the light rays that can reach or emanate from the endpoints of Alice's Gaussian support, indicating the region along Bob's trajectory at which Bob can send or receive signals  from Alice (via the coupling with the massless scalar field). We divide the regions along Bob's trajectory into four parts, shown in Fig.~\ref{fig:FS signalling}. Regions (i) and (iii) do not allow signalling between both detectors: this is expected from the fact that the commutator of the (proper-time derivative of the) field has support only on the lightlike region. Region (ii) is where Bob can signal to Alice when the Gaussian support of Bob's detector intersects the orange lines, whereas region (iv) is where Alice can signal to Bob. Notice that region (iv) is wider than region (ii).

For a given choice of Alice and Bob's detector parameters, the signalling region between Alice and Bob mediated by the detector-field interaction can be quantified by a single parameter $\delta$ and the PG coordinates. 
As depicted in Fig. \ref{fig:FS signalling}(a), we first find the constant $t\ts{PG}$ slice that crosses Alice's Gaussian peak $\tau\ts{A,0}$. This is simply given by $t\ts{PG}=\tau\ts{A,0}$. Bob is stationary at the  radial coordinate\footnote{Recall that in PG coordinates, the spatial slices are flat. Thus the coordinate separation between two radial coordinates $r_i,r_j$ is equivalently given by the proper separation $ \Delta r = |r_i-r_j| = d(r_i,r_j)$. 
This is not true for Schwarzschild coordinates.} $r\ts{B}=r\ts{s}+d(r\ts{A},r\ts{s})+d\ts{AB}$ and we suppose that Bob has the freedom to decide when to switch the detector on (for fixed width $\sigma$). 
The parameter $\delta$ gives a measure of time delay of Bob's switching away from the $t\ts{PG}=\tau\ts{A,0}$ line (the constant-$t\ts{PG}$ slice that matches Alice's Gaussian peak) and it is given by
\begin{align}
    \delta 
    &\coloneqq 
    \tau\ts{B,0}- \sqrt{ f(r\ts{B}) } 
    \bigg[\tau\ts{A,0}
        -2 r\ts{s} \sqrt{ \frac{r\ts{B}}{r\ts{s} } } - r\ts{s} \ln \frac{ \sqrt{ r\ts{B}/ r\ts{s} }-1 }{ \sqrt{ r\ts{B}/ r\ts{s} } +1 }
    \bigg]
    \label{eq:delta}
\end{align}
Note that $\delta>0$ if the Gaussian peak $\tau\ts{B,0}$ is located in the future of the constant $t\ts{PG}=\tau\ts{A,0} $ line.

\begin{figure*}[tp]
    \centering
    \includegraphics[scale=0.725,trim=0 0 0 0]{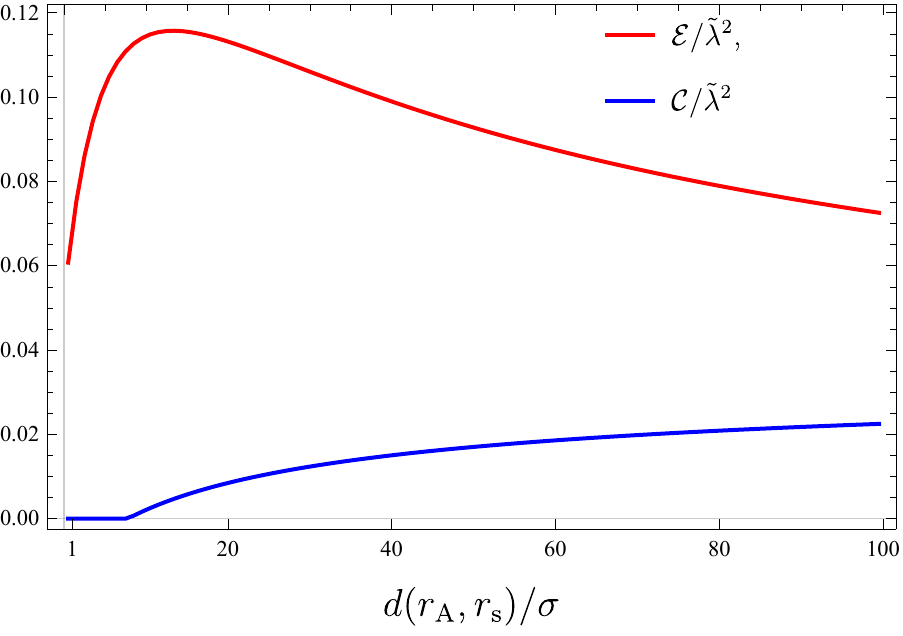}
    \includegraphics[scale=0.910,trim=0 +0.3cm 0 0]{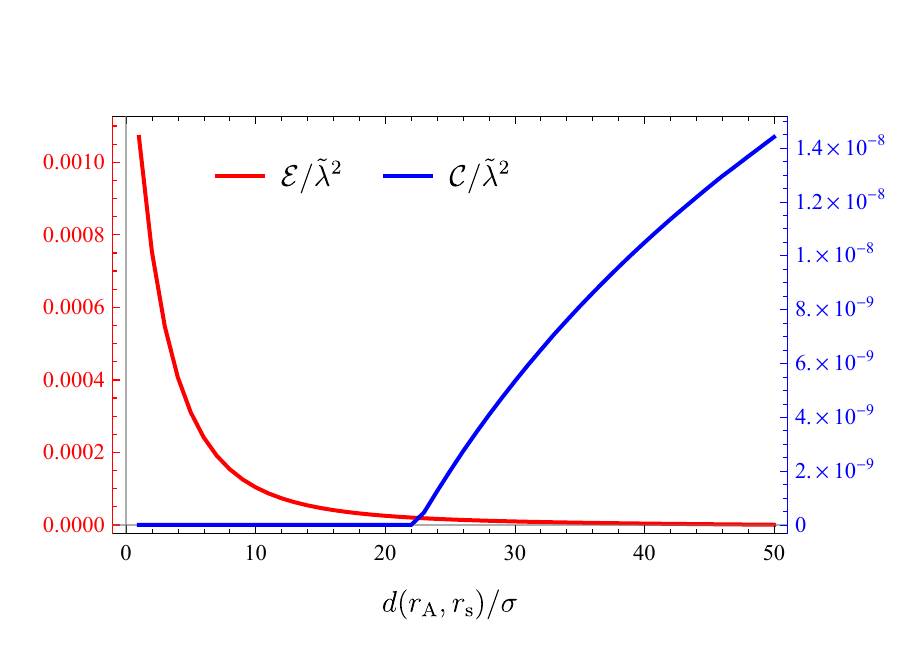}
    \caption{~Concurrence and signalling estimator as a function of proper distance from the horizon for FS scenario. \textbf{Left:} $\Omega \sigma=2, M/\sigma=5, d\ts{AB}=2\sigma$, as done in Fig.~\ref{fig:Concurrence FSvsSS}.  \textbf{Right:} $M/\sigma=5, \Omega \sigma=5, d\ts{AB}/\sigma=5$.  Here we chose $\delta/\sigma=1$ so that two detectors cannot communicate at large $d(r\ts{A}, r\ts{s})$.}
    \label{fig:Concurrence and signalling}
\end{figure*}

The signalling estimator between the two detectors as a function of $\delta/\sigma$ is given in Fig. \ref{fig:FS signalling}(b); we see that it is not   an entanglement monotone, as expected. 
By comparing with Fig.~\ref{fig:FS signalling}(a), we see that as $\delta $ varies from regions (i)-(iv), we see that the communication between Alice and Bob can be precisely captured using the signalling estimator $\mathcal{E}$ (or rather the absolute value $|\mathcal{E}|$). Indeed the signalling estimator $|\mathcal{E}|$ is very sharp and narrow in region (ii) because the support of Bob's Gaussian switching crosses the small region where Bob can send signals to Alice via the field. Region (iv), where Bob can receive signals from Alice, is very wide compared to (ii), and this manifests in $|\mathcal{E}|\neq 0$ for a wide range of $\delta$.  Regions (i) and (iii) have vanishing $|\mathcal{E}|$ because Bob is outside the signalling region. In region (v) $\delta$ is large enough that Bob is again causally disconnected from Alice.

We can now clarify whether the harvesting protocol we studied earlier is communication-assisted or not. By computing the signalling estimator for the results in Fig.~\ref{fig:Concurrence FSvsSS}, we conclude that indeed the harvesting protocol \textit{is} communication-assisted because $|\mathcal{E}|\neq 0$, as we show in the left plot of Fig.~\ref{fig:Concurrence and signalling}. In the language of Fig.~\ref{fig:FS signalling}, this also means that for $d\ts{AB}=2\sigma$, region (iii) is so small that it cannot completely contain the Gaussian support of Bob; hence the commutator cannot vanish as $\delta$ increases from (ii) to (iv). 

In view of this, we can ask whether the protocol still allows for entanglement harvesting with free-falling Alice when Bob is causally disconnected from Alice; the answer is yes, as we show in the right plot of Fig.~\ref{fig:Concurrence and signalling}. In this case we have chosen the setup parameters and manipulated the time-delay parameter $\delta$ such that the detectors are causally disconnected.  Indeed, we see that the two detectors can still have nonzero concurrence\footnote{Mutual information is generically nonzero everywhere except possibly when Alice approaches the curvature singularity, so nonzero concurrence is a more relevant and decisive measure.} purely from vacuum entanglement harvesting since $|\mathcal{E}|\approx 0$. Conversely, from the left diagram in  Fig.~\ref{fig:Concurrence and signalling} we also see that a nonzero communication channel mediated by the field ($|\mathcal{E}| >  0$) does not guarantee that two detectors can harvest entanglement.
This is analogous to the situation when a quantum communication channel is entanglement-breaking.

\begin{figure*}[tp]
\begin{center}

    \begin{tabular}{c}
    \begin{minipage}[c]{0.4\linewidth}
    \begin{center}
        \includegraphics[clip,width=6.3cm]{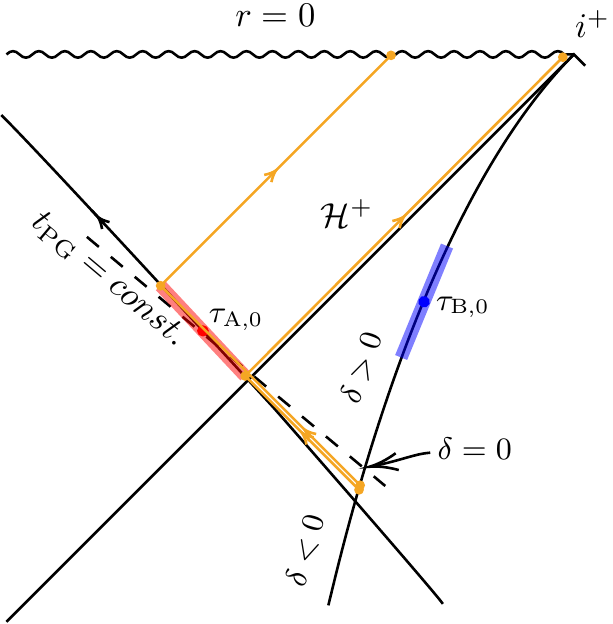}
        \hspace{1.6cm}(a)
    \end{center}
    \end{minipage}

    \begin{minipage}[c]{0.5\linewidth}
    \begin{center}
        \includegraphics[clip,width=\textwidth]{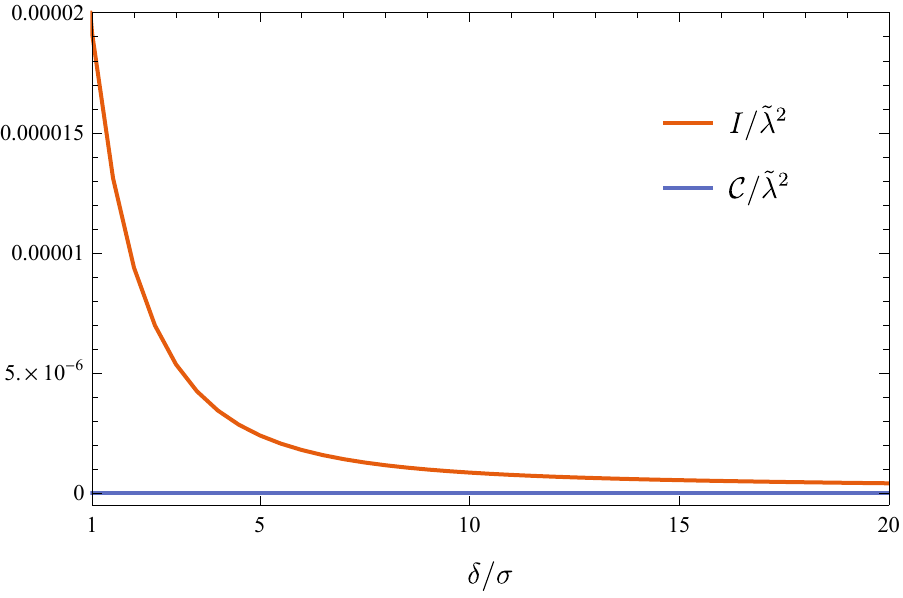}
        \hspace{1.6cm}(b)
    \end{center}
    \end{minipage}

\end{tabular}
\caption{~Schematics of the FS scenario when Alice's strong support is completely contained in the black hole interior. Here we chose $\Omega \sigma=2, M/\sigma=10, d(r\ts{A},0)/\sigma=14$, and $d(r\ts{B},r\ts{s})/\sigma=7$. 
Alice turns her detector on after she enters the black hole while Bob is staying outside. 
For this choice of parameters, mutual information harvested between the causally disconnected detectors are still manifestly nonzero, although the concurrence is zero in this case.}
\label{fig:causally disconnected mutual info}
\end{center}
\end{figure*}

Last but not least, we consider the final FS scenario when Alice's Gaussian support is completely contained inside the black hole, as shown in Fig.~\ref{fig:concurrence inside}. Here the crucial point is that Alice can never communicate to Bob because her causal future is completely contained in the black hole interior, although there are values of $\delta<0$ where Bob can still signal to Alice. Since Bob is static in the exterior, it is necessary that $d\ts{AB}$ be big enough for Alice's strong support to be completely inside the black hole and the black hole mass also needs to be big enough relative to the Gaussian width for the whole support to fit inside. An example that fits these requirements is given by $\Omega\sigma=2, M/\sigma=10, d(r\ts{A},0)/\sigma=14$, and $d(r\ts{B},r\ts{s})/\sigma=7$. For this choice, the concurrence is zero but the mutual information is still nonzero, as seen from Fig.~\ref{fig:causally disconnected mutual info}(b). 

While in Fig.~\ref{fig:causally disconnected mutual info}(b) the concurrence is zero, we have indirect numerical evidence that with large enough $\Omega$ harvesting nonzero concurrence across the horizon is possible. This is based on numerical evidence that
 $|\mathcal M|-\sqrt{\mathcal{L}\ts{AA}\mathcal L\ts{BB}}$ grows steadily with $\Omega$. 
 Unfortunately, due to highly oscillatory phase in the nonlocal matrix element $\mathcal{M}$, we are unable to plot the results for $\Omega\sigma\gtrsim 4.6$.  
 This numerical trend is commensurate with similar ones observed in flat space \cite{pozas2015harvesting}, an expanding (de Sitter) universe \cite{Steeg2009}, and the BTZ black hole \cite{henderson2018harvesting}.
 In the latter study  \cite{henderson2018harvesting}, once concurrence is nonzero for sufficiently large $\Omega$ (in units of $\sigma^{-1}$), it remains nonzero as $\Omega$ increases and tends to zero in the limit $\Omega\sigma\to\infty$. This is because on general grounds local noise $\mathcal{L}_{jj}$ decreases and $|\mathcal{M}|$ increases as $\Omega$ increases, although the behavior of $|\mathcal{M}|$ is in general not monotonic. 
 
 Thus, despite the horizon cutting off  causal connection from Alice to Bob, we expect the harvesting protocol can still extract entanglement from the vacuum with suitably chosen energy gap of the detector (and optimizing over other parameters). Note that mutual information harvesting is generically nonzero, as shown in Fig.~\ref{fig:causally disconnected mutual info}(b).

In Fig.~\ref{fig:causally disconnected mutual info}, the mutual information harvested decreases with increasing $\delta$. Thus it seems that the harvesting protocol depends on how late Bob's detector is turned on even though the proper separation between the two detectors are the same. This can be traced to the fact that while the Wightman two-point functions are stationary with respect to coordinate $t$ (Schwarzschild or PG), it is \textit{not} stationary with respect to the respective proper times $\tau\ts{A}$ and $\tau\ts{B}$, i.e., $\mathcal{A}_\alpha(\tau\ts{A},\tau\ts{B})\neq \mathcal{A}_\alpha(\tau\ts{A}-\tau\ts{B})$, since there is relative gravitational and kinematical redshift between the two trajectories. This is true even for two static detectors at two different radii, and hence is also true for any FS scenarios where the two trajectories do not share the same proper time parametrization. In that sense, the harvesting protocol is sensitive to the relative proper time delay between both detectors, i.e., making one of the detectors turn on   much later in  the future generically decreases the mutual information and concurrence between them.\footnote{Generically this is because by varying $\delta$ there will be a value $\delta_\text{max}$ where it attains a maximum for both concurrence and mutual information, since the same reasoning should hold if Bob is turned on ``too early'' in the asymptotic past. For derivative coupling, the lightlike support of the commutator implies that switching on too early also disables communication.}

We comment on the implementation of the harvesting protocol when Alice is inside the black hole. Strictly speaking, since Alice and Bob are causally disconnected by the horizon, there is no physical procedure for checking the entanglement by themselves. This is because neither party can collect the other party's detectors and perform state tomography of the joint system. For this particular scenario, we can follow similar principle as outlined in e.g. \cite{Steeg2009}: essentially, one has to consider a third party, say Charlie, who follows a trajectory that is contained in the causal futures of \textit{both} Alice and Bob's detectors. Charlie will then collect information from both parties and perform state tomography on their behalf. Note that Charlie also needs to fall \textit{inside} the black hole, since Alice's causal future is contained in the black hole interior. In contrast, when both detectors are outside, Alice and Bob can simply reconvene after the interactions have been turned off, although they can also employ a third party to do the joint state tomography.

\subsection{Alice and Bob free-falling (FF)}

We close the section by briefly analyzing the FF scenario, i.e. when both Alice and Bob are free-falling towards the black hole. In this paper we only consider the free-falling trajectory initially at rest at spatial infinity since the adapted coordinate system is precisely the PG coordinates. Thus for this FF setup, Alice and Bob follow the same  timelike trajectory but with different switching peaks. This simpler setup has the advantage that the derivative-coupling Wightman functions are expressible in relatively simple terms, since the radially infalling geodesics are relatively straightforward to implement.  Note that due to the derivative coupling with the field, the support of $\comm$ is lightlike. Thus being along the same timelike trajectory does not guarantee field-mediated communication between them.

\begin{figure*}[t]
\begin{center}
    \begin{tabular}{c}
    \begin{minipage}[c]{0.35\linewidth}
    \begin{center}
        \includegraphics[clip,width=\textwidth]{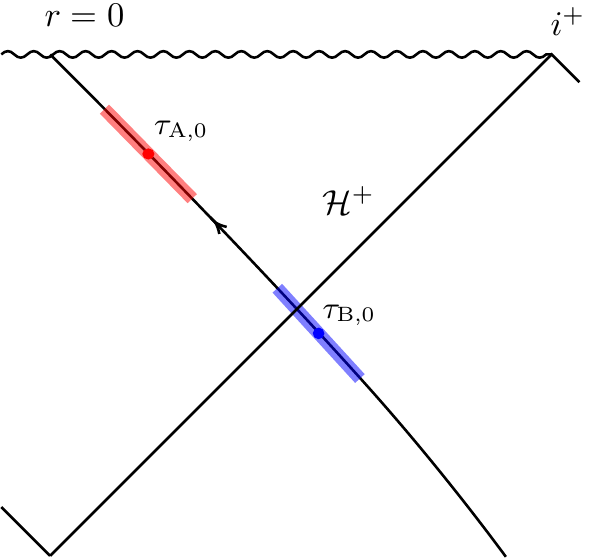}
        \hspace{1.6cm}(a)
    \end{center}
    \end{minipage}
    
    \begin{minipage}[c]{0.6\linewidth}
    \begin{center}
        \includegraphics[clip,width=\textwidth]{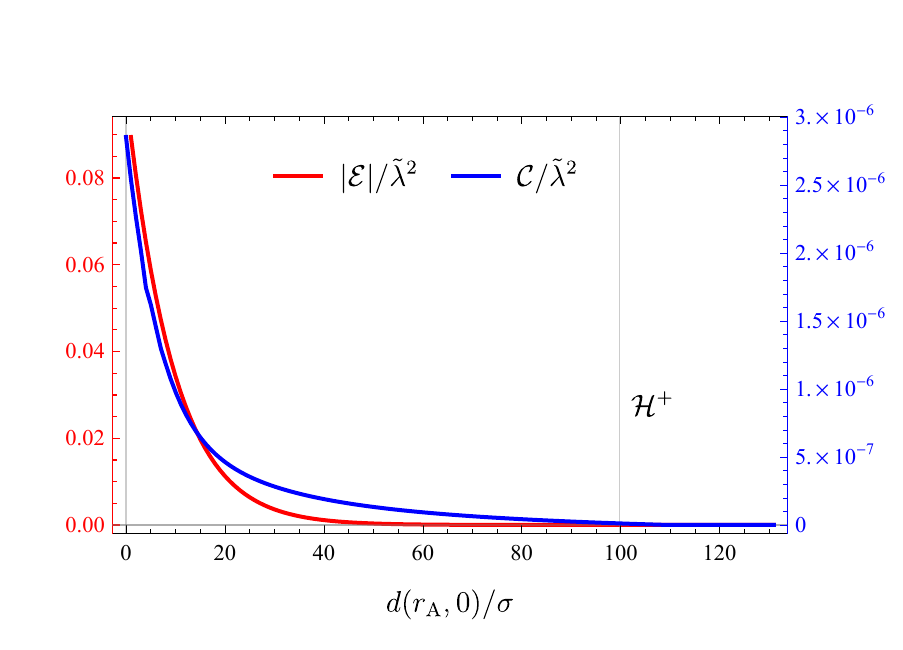}
        \hspace{1.6cm}(b)
    \end{center}
    \end{minipage}
    \end{tabular}
    \caption{~ Schematics of two detectors in the FF scenario.  Here we set $\Omega \sigma=5,M/\sigma=50, d\ts{AB}/\sigma=5$. (a) The Penrose diagram of freely falling detectors, with Alice infalling first towards the singularity followed by Bob. The horizontal axis is Alice's proper distance from the singularity, and the event horizon is at $r\ts{s}/\sigma=2 M/\sigma=100$.  Alice cannot signal to Bob via the coupling to the field but can receive signals from Bob when they are close to the singularity. (b) The modulus of signalling estimator $|\mathcal{E}|/\tilde{\lambda}^2$ (red) and the concurrence (blue) between the two free-falling observers.
    }
    \label{fig:concurrence inside}
\end{center}
\end{figure*}

As shown in Fig.~\ref{fig:concurrence inside}(a), without loss of generality we set Bob's detector to switch on earlier than Alice's. This  simulates the effect of Alice free-falling ahead of Bob. The effective proper separation  $d\ts{AB}$ between the detectors is fixed with respect to their Gaussian peaks. That is, given Alice's Gaussian peak at $\tau\ts{A,0}$ and a fixed proper distance $d\ts{AB}$, we can work out at what proper time $\tau\ts{B,0}$ Bob's Gaussian peak should be. More explicitly, given Alice's (effective) proper distance from the horizon $d\ts{A}\coloneqq d(r\ts{A},0)$, their Gaussian peaks can be written purely in geometric terms as 
\begin{align}
    \tau\ts{A,0} &=-\frac{d\ts{A} }{3} \sqrt{\frac{2d\ts{A}}{M} }\,,\\
    \tau\ts{B,0} &=-\frac{d\ts{A}+d\ts{AB}}{3} \sqrt{\frac{2(d\ts{A} + d\ts{AB}) }{M} }\,.
\end{align}
We can  study the harvesting protocol in a manner  analogous to the SS or FS scenarios for any choice of $d\ts{A}$ and $d\ts{AB}$. Fig. \ref{fig:concurrence inside}(b) depicts the signalling estimator and the concurrence between two freely falling detectors when $\Omega \sigma=5,M/\sigma=50, d\ts{AB}/\sigma=5$.


We immediately notice from  Fig.~\ref{fig:concurrence inside}(b) that the previously known entanglement shadow near the horizon \cite{henderson2018harvesting, Tjoa2020vaidya} is absent; the freely falling detectors can harvest entanglement from the interior and exterior of the black hole. 
First, note that in the SS scenario, the closer the detectors are to the horizon, the lower the value of $|\mathcal{M}|-\sqrt{\mathcal{L}\ts{AA}\mathcal{L}\ts{BB}}$ as the noise term $\mathcal{L}_{jj}$ dominates the nonlocal term $|\mathcal{M}|$ \cite{henderson2018harvesting,Tjoa2020vaidya}. On general grounds, it can be shown that for  fixed proper separation $d\ts{AB}$,  $|\mathcal{M}|$ remains finite as Alice (and hence Bob) is brought closer to the horizon (in fact vanishes in this limit), whilst $\mathcal{L}\ts{AA} > |\mathcal{M}|$ (for sufficiently small $d\ts{AB}$ we have $ \mathcal{L}\ts{AA}\sim \mathcal{L}\ts{BB}$). This behavior is a generic result of the fact that static detectors cannot remain static at the horizon, manifest as an   infinite gravitational redshift factor.  In contrast, the equivalence principle requires that free-falling observers (or detectors) experience nothing peculiar across the horizon. Free-falling detectors do not experience divergent gravitational redshift at the horizon, and so in generic FF situations  the entanglement shadow is absent as both detectors cross the horizon along a radial geodesic.

Moreover, Fig.~\ref{fig:concurrence inside}(b) shows that the amount of harvested entanglement (in blue) increases as the detectors reach the singularity. 
At early stages after horizon crossing, $|\mathcal{E}|$ is small and most of the entanglement harvested is from the vacuum. As the singularity is approached, we can (using analogous ray-tracing analysis   from previous subsections) see that  as the singularity is approached, Bob becomes able to signal to Alice, which is  manifest as increasing nonzero $|\mathcal{E}|$ (in red).  Therefore, up until the point at which Alice's Gaussian tail  approaches the singularity (beyond which we cannot make any conclusions at a semiclassical level), entanglement harvesting  becomes increasingly communication-assisted for this particular setup.

Finally we make a comment regarding the generality of this FF scenario. 
There are extra complications in both analytic and numerical evaluation of the detectors' reduced density matrix if we consider generalized free-falling coordinates associated with free-falling observers initially at rest at finite radial coordinate $r_0$ (see \cite{Martel2001generalPG} for the generalized coordinates adapted to different free-falling observers). This modified scenario is very useful in principle because we can minimize the effect of relative velocities of the detectors when they are very close to the black hole or even inside: our current setup necessarily requires that Alice has non-negligible (possibly relativistic) velocities that may by itself suppress the quality of the harvesting protocol. We leave this finite-radius free-falling scenario for future work.

\section{Conclusion}
\label{sec: conclusion}

We have analyzed the harvesting protocol for both entanglement and mutual information    in (1+1)-dimensional Schwarzschild spacetime involving a combination of static and freely falling detectors. Employing the derivative coupling UDW model, which removes the IR ambiguity and mimics the Hadamard short-distance property of massless scalar field in $(3+1)$-dimensional case, 
we considered three possible scenarios for comparison based on  Alice and Bob's detector trajectories: static-static (SS), free-falling-static (FS), and both free-falling (FF).

First, we found that within perturbation theory, the concurrence and the mutual information for the FS scenario is always less than that of the SS scenario.  We identified the origin of this relative inefficiency to be largely kinematic, insofar as the relative velocity between the detectors is the main cause the degradation of correlations. This calculation is enabled by the generalization of relative velocities in curved spacetimes \cite{bolos2007intrinsic}. A comparison of the FS scenario in the Boulware vacuum to the case of two detectors with the same relative velocity in Minkowski spacetime confirms this. Moreover, the so-called entanglement shadow (or `death-zone') near the horizon for the FS case is much larger than that of the SS one found previously \cite{henderson2018harvesting,robbins2020entanglement,Tjoa2020vaidya}.  It is an interesting question as to whether or not there are contributions to the amount of correlations harvested that can be attributed to ``truly intrinsic" properties of the gravitational field that the concept of generalized relative velocities exclude. Our results, at least for sufficiently large distance from the black hole, suggest that this would occur only for near-horizon regimes or within black hole interior.

Second, we investigated to what extent the harvesting protocol is communication-assisted by examining the signalling estimator $\mathcal{E}$ in \eqref{estimator}
and studying the causal relation between the detectors. We were able to show that in the FS scenario the detectors can harvest entanglement purely from the QFT vacuum. In addition, when the detectors are causally disconnected by the event horizon, we show that they can still harvest correlations as well directly from the field vacuum. Note that due to the derivative coupling, timelike separated detectors cannot communicate via the field since the support of commutator of the proper time derivative of the field is lightlike.

Finally, we analyzed a simple scenario involving both detectors free-falling towards the black hole, where both detectors follow the same geodesic but they turn on the detectors in different times. We find that the entanglement shadow is absent; this is in accord with the equivalence principle, {and can be attributed to the fact that nothing peculiar should happen during horizon crossing for
detectors on inertial trajectories.  This is in contrast to static detectors, which require increasing local acceleration as the horizon is approached, and which cannot remain static 
at the horizon.} 

There are several interesting future directions that can be pursued based on our results.
It would be most interesting to confirm cross-horizon entanglement harvesting. This might be done by
considering free-falling detectors at finite radii in the bulk geometry, employing adapted coordinates outlined in \cite{Martel2001generalPG}. Although this would minimize the relative velocities via the initial conditions imposed on the trajectories, it would be at the price of greater complexity in the evaluation of the density matrix elements of the detectors (especially numerically). 
Second, it would be interesting to consider the free-falling detector scenario when the black hole has different interior structure (such as regular black holes without a curvature singularity \cite{Hayward2006regular,Frolov2016Hayward}). It would also be interesting to see how these results can be reconciled with techniques based on master equations (see, e.g., \cite{Moustos2014master,Scully2018free-fall} and references therein). 
We leave these interesting directions for future work.

\section*{Acknowledgment}

The authors thank Eduardo Mart\'in-Mart\'inez and Sayan Gangopadhyay for useful discussions.
K.G-Y. acknowledges the support from Keio University Global Fellowship and 
E.T. acknowledges the support from the Mike and Ophelia Lazaridis Fellowship during this work. This work was supported in part by the Natural Sciences and Engineering Research Council of Canada and by Asian Office of Aerospace Research and Development Grant No. FA2386-19-1-4077.

\appendix
\section{The geodesic equations and the null coordinates}
\label{appendix: geodesic}

In this Appendix we calculate the solution to the radial geodesic equation in Schwarzschild coordinates, with the goal of expressing the null coordinates $u,v,U,V$ that appear in the definition of derivative coupling Wightman functions as functions of the free-falling detector's proper time.

Consider a freely falling observer who is initially at rest at $r_0>2M$. 
The geodesic equations for a radially infalling observer are given by \cite{carroll2019spacetime}
\begin{align}
    \ddot{t}&=-\dfrac{2M}{r^2 f(r)} \dot{t} \dot{r}\,, \\
    \ddot{r}&=-\dfrac{f(r)M}{r^2}\dot{t}^2 + \dfrac{M}{r^2 f(r)}\dot{r}^2\,,
    \label{eq:geodesic eq}
\end{align}
where the dotted derivative refers to the derivative with respect to proper time. The differential equation is also supplemented with a constraint for a massive test particle, namely that the   geodesic is timelike:
\begin{align}
    -1=g_{\mu \nu} \dot{x}^\mu \dot{x}^\nu 
    =-f(r) \dot{t}^2 + \dfrac{1}{f(r)} \dot{r}^2, \label{eq:constraint test particle}
\end{align}
where $\dot{x}^\mu=\dd x^\mu/\dd\tau$ is the four-velocity. By multiplying $M/r^2$ on both sides of \eqref{eq:constraint test particle} and substituting into \eqref{eq:geodesic eq}, we get
\begin{align}
    \ddot{r}=-\dfrac{M}{r^2}\,. \label{eq:acceleration}
\end{align}
By integrating this with the initial conditions, $r(\tau_0)=r_0,~\dot{r}(\tau_0)=0$,
where $\tau=\tau_0$ is the initial proper time, we obtain
\begin{align}
    \dot{r}= -\sqrt{ \dfrac{r\ts{s}}{r} - \dfrac{r\ts{s}}{r_0} } \,.
\end{align}
Assuming that the observer starts from $r_0=\infty$, the geodesic equations reduce to
\begin{align}
    \dot{t}&=
    \dfrac{1}{ 1-r\ts{s}/r }\,, \label{eq:t dot}\\
    \dot{r}&= -\sqrt{ r\ts{s}/r }\,,\label{eq:r dot}
\end{align}
and so $r$ can be obtained as \cite{Aubry2014derivative}
\begin{align}
    r(\tau)&=r\ts{s} 
    \kako{
        \dfrac{\tau}{\tau\ts{s}}
    }^{\frac{2}{3}}, \label{eq:geodesic r solution}
\end{align}
where $\tau\ts{s}=-4M/3$ is the horizon-crossing time. 
Note that $\tau\in (-\infty, 0)$, and the observer reaches the singularity as $\tau\to 0^-$. 

Let us rewrite the geodesic equations in terms of the two null coordinates $v,u=t\ts{s}\pm r_\star$ and the Kruskal-Szekeres null coordinates $V= 2r\ts{s} e^{v/(2r\ts{s})}$ and $U = -2r\ts{s} e^{-u/(2r\ts{s})}$. By using \eqref{eq:t dot} and \eqref{eq:r dot}, we get
\begin{align}
    \dot{v}&= \dfrac{1}{ 1 + \sqrt{ r\ts{s}/r } }\,, \\
    \dot{u}&= \dfrac{1}{ 1 - \sqrt{ r\ts{s}/r } }\,, \\
    \dot{V}&=
    \dfrac{V}{2r\ts{s}}
    \dfrac{1}{ 1 + \sqrt{ r\ts{s}/r } }\,, \\
    \dot{U}&=
    -\dfrac{U}{2r\ts{s}}
    \dfrac{1}{ 1 - \sqrt{ r\ts{s}/r } }\,,
\end{align}
for $r>0$. Note that due to the local nature of the differential equations, the geodesic equations expressed in terms of these null coordinates extend into the black hole interior, in contrast to the Schwarzschild coordinate version which is only valid in the exterior region. Substituting \eqref{eq:geodesic r solution} into these differential equations, we can obtain closed form expressions for $u,v,U,V$ as functions of infalling $\tau$,
\begin{align}
    v(\tau)
        &=
        \tau
        -2r\ts{s} 
        x^{ \frac{1}{3} }(\tau)
        +
        r\ts{s}
        x^{ \frac{2}{3} }(\tau)
        +2r\ts{s}
        \ln 
        \kagikako{
            1 + 
            x^{ \frac{1}{3} }(\tau)
        }, \\
        u(\tau)
        &=
        \tau
        -2r\ts{s} 
        x^{ \frac{1}{3} }(\tau)
        -
        r\ts{s}
        x^{ \frac{2}{3} }(\tau)
        -2r\ts{s}
        \ln 
        \kagikako{
            -1 + 
            x^{ \frac{1}{3} }(\tau)
        }, \\
   V(\tau)
   &=
   2r\ts{s}
   e^{ \frac{\tau}{2r\ts{s}} }
   \exp
   \kagikako{
            - 
            x^{ \frac{1}{3} }(\tau)
            +\dfrac{1}{2}
            x^{ \frac{2}{3} }(\tau)
        }
        \kako{
                1+
                x^{\frac13}(\tau)
            }
   , \\
   U(\tau)
   &=
   2r\ts{s}
   e^{ - \frac{\tau}{2r\ts{s}} }
   \exp
    \kagikako{
            x^{ \frac{1}{3} }(\tau)
            +
            \dfrac12
            x^{ \frac{2}{3} }(\tau)
        }
        \kako{
                1-
                x^{\frac13}(\tau)
            }\,,
\end{align}
where $x(\tau)=\tau/\tau\ts{s}$. 
One can check that at the singularity $r=0$, we get $UV=4r\ts{s}^2$, as expected. Furthermore, these null coordinates take the extended values in the relevant regions; for instance, while the coordinate transformations for $U,V$ are given for $U<0,V>0$, the solutions $U(\tau),V(\tau)$ can take all real values and hence include the black hole interior.

\bibliography{FFref}

\end{document}